\documentclass[a4paper,fleqn]{article}
\usepackage[numbers,sort&compress]{natbib}
\usepackage{graphicx}
\usepackage{latexsym}
\usepackage{amsmath,amssymb}
\usepackage{dcolumn}
\usepackage{float}
\usepackage{caption}

\begin{document}
\captionsetup[figure]{labelfont={bf},labelformat={default},labelsep=space,name={Fig.}}

\title{{\bf Structure of geodesics in the regular Hayward black hole space-time}
\author{\normalsize Jian-Ping Hu$^{1}$,
       Yu Zhang$^{1}$\thanks{Corresponding Author(Y. Zhang): Email: zhangyu\_128@126.com}, Li-Li Shi$^{1}$,
        and Peng-Fei Duan$^{2,3}$ \\
   \normalsize \emph{$^{1}${Faculty of Science, Kunming University of Science and Technology, }}\\ \normalsize \emph {Kunming, Yunnan 650500, People's Republic of China}\\
   \normalsize \emph{$^{2}$Geological Resources and Geological Engineering Postdoctoral Programme, Kunming University of} \\ \normalsize \emph {Science and Technology, Kunming, Yunnan 650093, People's Republic of China}\\
   \normalsize \emph{$^{3}$City College, Kunming University of Science and Technology, Kunming, Yunnan 650051, }\\ \normalsize \emph{People's Republic of China}}}

\date{}
\maketitle \baselineskip 0.3in

{\bf Abstract} \ {The regular Hayward model describes a non-singular black hole space-time. By analyzing the behaviors of effective potential and solving the equation of orbital motion, we investigate the time-like and null geodesics in the regular Hayward black hole space-time. Through detailed analyses of corresponding effective potentials for massive particles and photons, all possible orbits are numerically simulated. The results show that there may exist four orbital types in the time-like geodesics structure: planetary orbits, circular orbits, escape orbits and absorbing orbits. In addition, when $\ell$, a convenient encoding of the central energy density $3/8\pi\ell^{2}$, is $0.6M$, and $b$ is $3.9512M$ as a specific value of angular momentum, escape orbits exist only under $b>3.9512M$. The precession direction is also associated with values of $b$. With $b=3.70M$ the bound orbits shift clockwise but counter-clockwise with $b=5.00M$ in the regular Hayward black hole space-time. We also find that the structure of null geodesics is simpler than that of time-like geodesics. There only exist three kinds of orbits (unstable circle orbits, escape orbits and absorbing orbits).

}


\section{Introduction}
The regular Hayward black hole model was presented by S. A. Hayward\cite{hayward2006angular} in 2006. It is non-singular, just like the Bardeen black hole\cite{bardeen1968non}. The regular Hayward black hole consists of a compact space-time region of trapped surfaces, with inner and outer boundaries which join circularly as a single smooth trapping horizon\cite{hayward2006angular}. The lapse function\cite{zhou2012geodesic} of the regular Hayward black hole space-time is $f(r)=1-\frac{2Mr^{2}}{r^{3}+2Ml^{2}}$, a new parameter $\ell$ (a convenient encoding of the central energy density $3/8\pi\ell^{2}$)\cite{hayward2006angular} is introduced into the function. Many researchers have done a lot of work around the regular Hayward black hole\cite{lin2013quasinormal,abbas2014geodesic,amir2015rotating,amir2016collision}. For $\ell$ = 0, the metric of the regular Hayward black hole will revert to the metric of Schwarzschild black hole.

It is known that many gravitation effects exist near a black hole, such as gravitational time-delay\cite{bozza2004time}, bending of light\cite{delorenci2001dyadosphere,virbhadra2008relativistic}, gravitational red-shift\cite{ross1971gravitational}, precession of planetary orbits\cite{greenberg1981apsidal} and quasinormal modes\cite{cardoso2003quasinormal,berti20009quasinormal,konoplya2011quasinormal}, etc. Investigating the geodesic structure \cite{chandrasekhar1984the,stuchlik1991null,cruz1994geodesic,beem1997stability,podolsky1999thestructure,breton2002geodesic,allison2003geodesic,kraniotis2004precise,stuchlk2004equatorial,hackmann2008geodesic,cardoso2009geodesic,abdujabbarov2010test,muller2011studying,halilsoy2013rindler,chakraborty2014inner-most,pradhan2015circular,zhang2015time-like,chandler2015geodesic,konoplya2017areeikonal,farrugia2017thermodynamic,azam2017geodesic,azam2017geodesic2} of the black hole space-time can help us understand these gravitational effects. In recent years, it has become an important and challenging topic in the astrophysical studies of black hole. In order to explain the structure of geodesics better, the researchers have done a lot of work. Different research methods have been presented, including the effective potential analysis method\cite{chen2010timelike,sheng2011time-like,li2014particle} and the phase plane method \cite{dean1999phase,zhang2014orbital,zhang2014orbital2,yi2006geodesics}. By analyzing the behavior of effective potential, Juhua Chen et al.\cite{chen2010timelike} investigated the geodesics of particles in Horava-Lifshitz space-time\cite{Petr2009Quantum}, and Sheng Zhou et al.\cite{sheng2011time-like} studied the time-like geodesic structure of the spherically symmetric black hole space-time\cite{heydarifard2009spherically}. By adopting the phase plane method, the time-like geodesic of spherical dilaton space-time\cite{frolov1987charged} were obtained by Yi Zeng et al.\cite{yi2006geodesics}. Their work has shown whether stable or unstable orbits exist is determined by angular momentum $b$, and which orbit the particles can move in depends on the energy levels $E^{2}$ of the particles.

Abbas, G. and Sabiullah, U.\cite{abbas2014geodesic} investigated the geodesic of the regular Hayward black hole. First they obtained the analytic solution for time-like and null geodesic in radial motion ($J=0$), i.e., the relationship between $r$ and $t$ (or proper time $\tau$). They also got the effective potential for non-radial motion ($J\neq0$) and radial motion ($J=0$). And the influence of angular momentum $J$ and energy $E$ on the effective potential of circular motion was also described. But they did not study and analyze how massive particles and photons move in this black hole space-time, the stability of motion and the different types of motion. For this reason, the orbital motion of massive particles and photons for regular Hayward black hole space-time is further analyzed in this paper. Although Abbas, G., Sabiullah, U. and we both investigate geodesic of the regular Hayward black hole space-time, our research is very different from theirs. By getting the equation between $R$ ($R = 1/r$) and azimuth angle $\phi$, we use numerical methods to directly plot the orbits of massive particles and photons at different energy levels. To know the influence of parameters on time-like geodesics, the behaviors of effective potential are analyzed in detail and all possible orbits numerically simulated. Furthermore, the structures of the time-like geodesic with different angular momentum $b$ are compared, and the influence of different energy levels $E^{2}$ on the same orbital type is discussed in null geodesic.

 The main purpose of this paper is to investigate the structure of geodesics by analyzing the behavior of effective potential and solving the equation of orbital motion in the regular Hayward black hole space-time. All possible orbits of test particles and photons are numerically simulated in the regular Hayward black hole space-time. The organization of the work is as follows: In Sect.2, we obtain the equation of orbital motion and define the effective potential. In Sect.3 and 4, we study the trajectories of the time-like and null geodesics in the regular Hayward black hole space-time in detail. Finally, the results and conclusions of our work are presented in last section.

\section{The equation of orbital motion and the effective potential}
\begin{figure}[H]
 \centering
    \includegraphics[angle=0, width=0.45\textwidth]{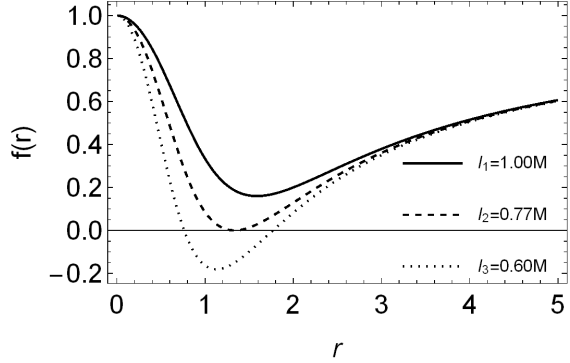}
     \vspace*{8pt}
 \caption{Horizons of the regular Hayward black hole space-time with different values of $\ell$ \label{v1} }
 \end{figure}

The metric of the regular Hayward black hole space-time is described as\cite{hayward2006angular}
\begin{eqnarray}
ds^{2}=-f(r)dt^{2}+\frac{1}{f(r)}dr^{2}+r^{2}d\theta^{2}+r^{2}\sin^{2}\theta d\phi^{2},
\label{e1}
\end{eqnarray}
\begin{eqnarray}
f(r)=1-\frac{2Mr^{2}}{r^{3}+2Ml^{2}},
\label{e2}
\end{eqnarray}
where $M$ is the mass of the regular Hayward black hole and $\ell$ is a convenient encoding of the central energy density $3/8\pi \ell^{2}$. Eq. (\ref{e2}) is the lapse function of the regular Hayward black hole space-time. Based on Eq. (\ref{e2}), three kinds of different space-time\cite{abbas2014geodesic} can be derived with no horizon (${\frac{\ell^{2}}{M^{2}}}>\frac{16}{27}$), one horizon(${\frac{\ell^{2}}{M^{2}}} = \frac{16}{27}$), and double horizons(${\frac{\ell^{2}}{M^{2}}} < \frac{16}{27}$), as shown in Fig. \ref{v1}.
As is known, the variation on the Euler-lagrange equations related to space-time metric can describe the geodesics. The corresponding Lagrangian according to the metric of the regular Hayward black hole can be written as
\begin{eqnarray} L=\frac{1}{2}m(\frac{ds}{d\tau})^2=-\frac{1}{2}mf(r)\dot{t}^2+\frac{1}{2}m\frac{1}{f(r)}\dot{r}^2+\frac{1}{2}mr^2\dot{\theta}^2+\frac{1}{2}mr^{2}\sin^{2}\theta\dot{\phi}^2.
 \label{e3}
\end{eqnarray}
In Eq. (\ref{e3}), $m$ is the mass of test particles, $\tau$ is the proper time and $\dot{t} = dt/d\tau$,\  $\dot{r} = dr/d\tau$, \ $\dot{\theta} = d\theta/d\tau$, \ $\dot{\phi} = d\phi/d\tau$. Without losing generality, we choose $\theta = \frac{\pi}{2}$, and Eq. (\ref{e3}) will become
\begin{eqnarray}
L=-\frac{1}{2}mf(r)\dot{t}^2+\frac{1}{2}m\frac{1}{f(r)}\dot{r}^2+\frac{1}{2}mr^{2}\dot{\phi}^2.
 \label{e4}
\end{eqnarray}
Thus, the Lagrangian equation is independent of $t$ and $\phi$. By using the Euler-Lagrangian differential equation, $\frac{d}{d\tau}(\frac{\partial L}{\partial \dot{x}^{\nu}})-\frac{\partial L}{\partial x^{\nu}}=0$, two more equations can be derived
\begin{eqnarray}
\frac{\partial L}{\partial t}=0 \Longrightarrow{-\frac{\partial L}{\partial \dot{t}}=\varepsilon=mf(r)\dot{t}}, \ \ \ \ \ \ \ \
\frac{\partial L}{\partial \phi}=0 \Longrightarrow{-\frac{\partial L}{\partial \dot{\phi}}=J=mr^{2}\dot{\phi}}.
\label{e5}
\end{eqnarray}
From Eq. (\ref{e5}), we obtain two constants: $\varepsilon$ (the total energy) and $J$ (the total angular momentum). Let $E=\frac{\varepsilon}{m}$ and $b=\frac{J}{m}$, and Eq. (\ref{e5}) can rewritten as
\begin{eqnarray}
\dot{t}=\frac{E}{f(r)}, \ \ \ \dot{\phi}=\frac{b}{r^{2}}.
\label{e6}
\end{eqnarray}
Then we can get the equation of orbital motion\cite{abbas2014geodesic}
\begin{eqnarray}
\dot{r}^{2}=E^{2}-(1-\frac{2Mr^{2}}{r^{3}+2Ml^{2}})(h+\frac{b^{2}}{r^{2}}).
\label{e7}
\end{eqnarray}
Defining the effective potential as
\begin{eqnarray}
 V_{eff}^{2}=(1-\frac{2Mr^{2}}{r^{3}+2Ml^{2}})(h+\frac{b^{2}}{r^{2}}).
\label{e8}
\end{eqnarray}

\section{The structure of time-like geodesics}
\subsection{The behavior of effective potential for time-like geodesics}
For massive particles, $h=1$, and the corresponding equation of effective potential should be written as
\begin{eqnarray}
 V_{eff}^{2}=(1-\frac{2Mr^{2}}{r^{3}+2Ml^{2}})(1+\frac{b^{2}}{r^{2}}).
\label{e9}
\end{eqnarray}

\begin{figure}[H]
\centering
    \includegraphics[angle=0, width=0.4\textwidth]{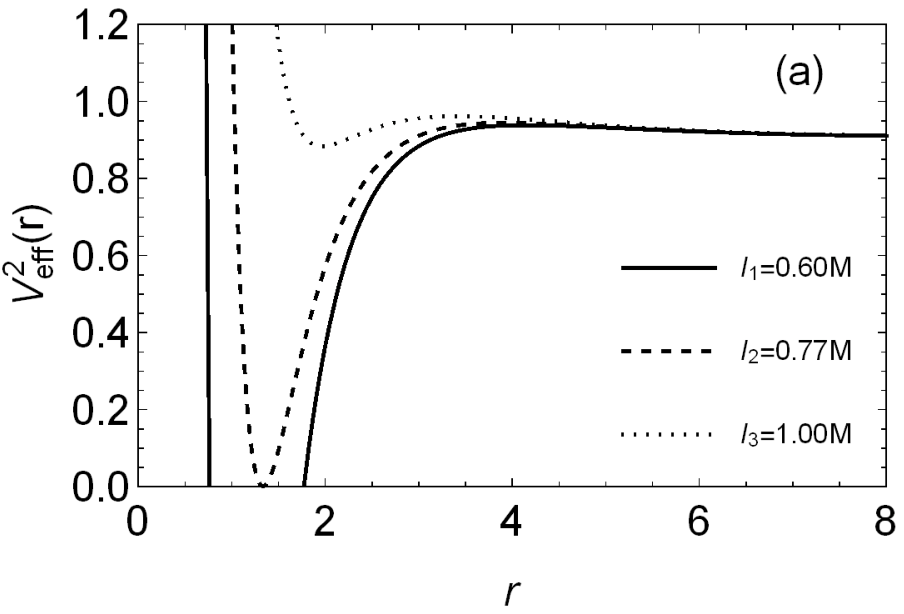}
    \ \ \ \  \includegraphics[angle=0, width=0.39\textwidth]{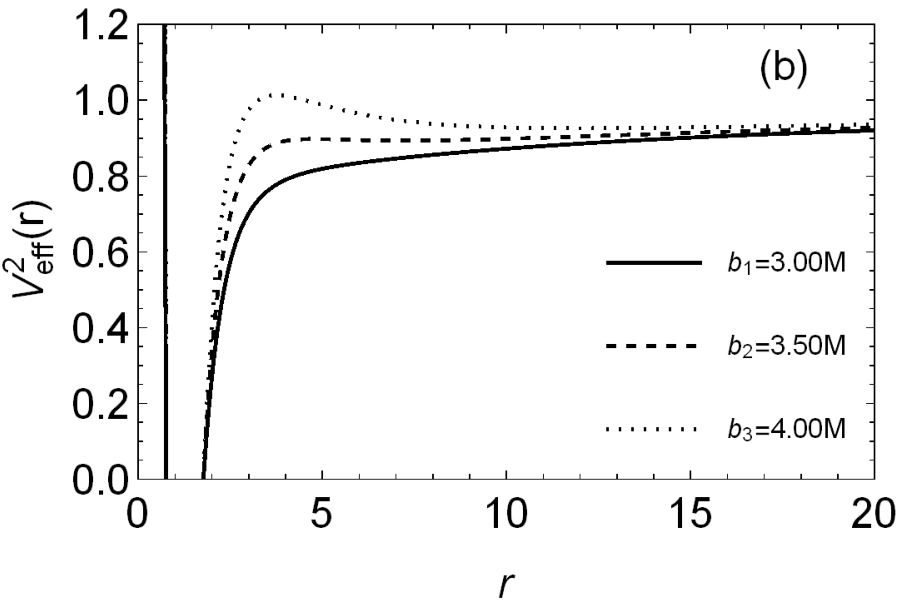}
     \vspace*{8pt}
\caption{The curves of effective potential with different values of parameters for fixed $b=3.70M$ (left) and fixed $\ell=0.60M$ (right). \label{v2} }
\end{figure}
\begin{figure}[H]
\centering
    \includegraphics[angle=0, width=0.39\textwidth]{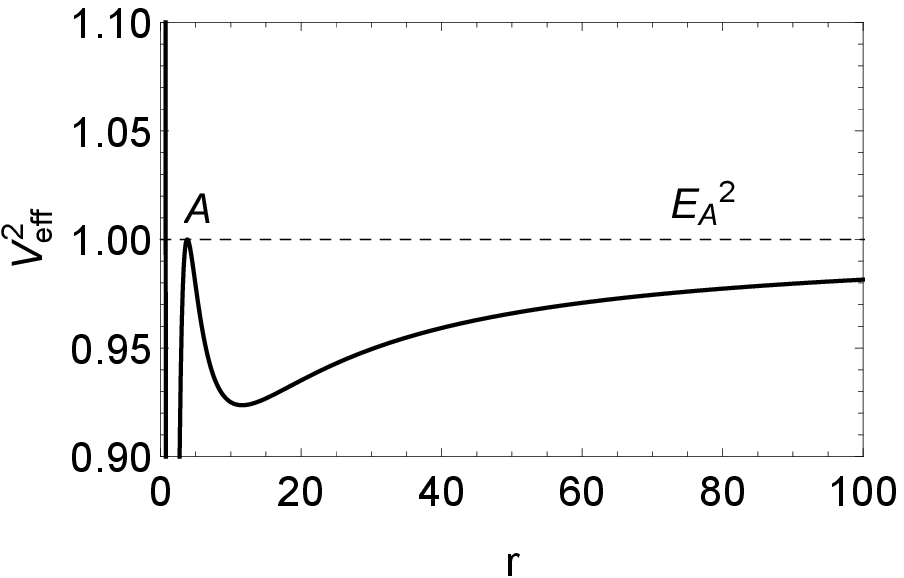}
     \includegraphics[angle=0, width=0.355\textwidth]{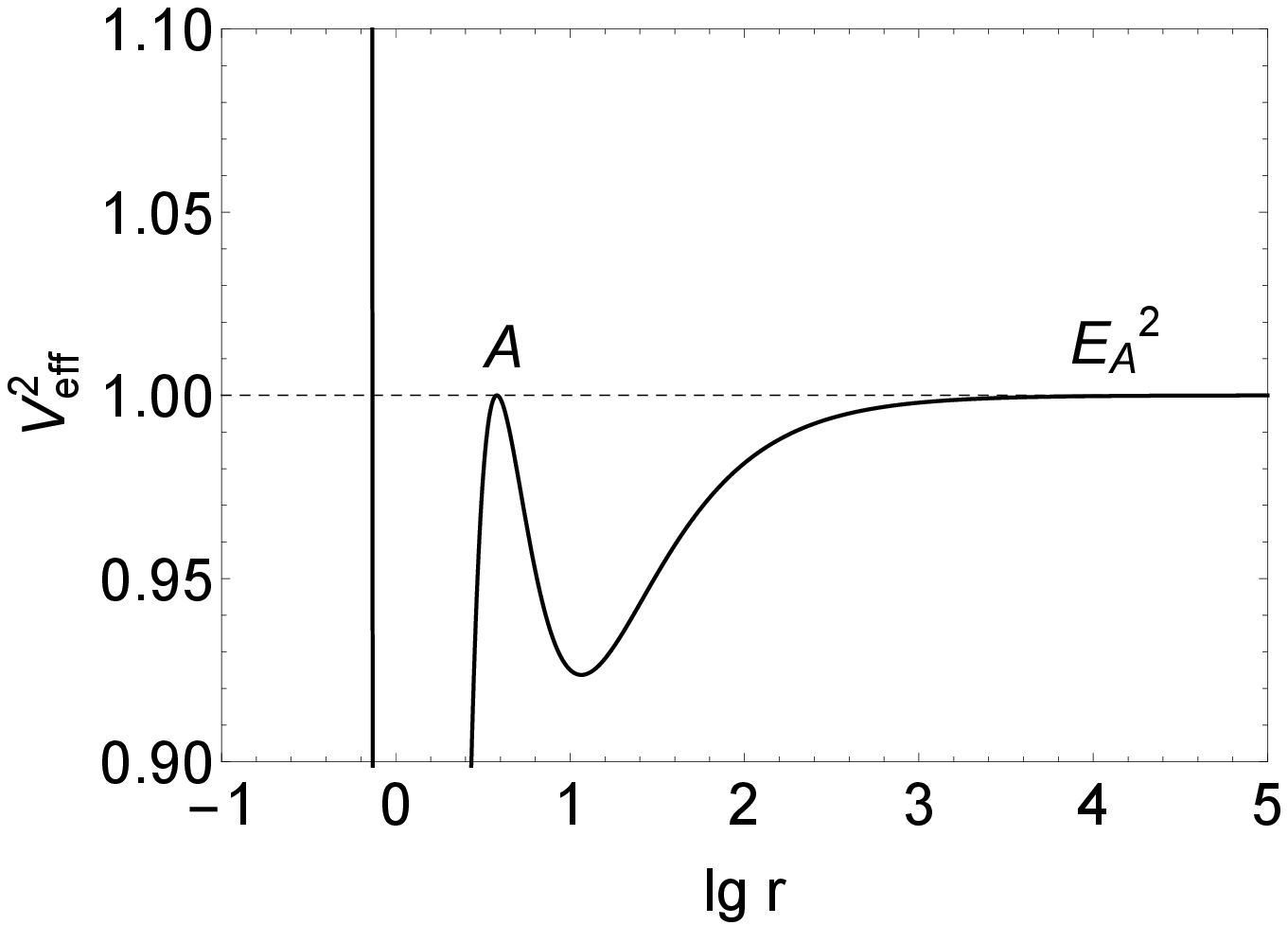}
     \vspace*{8pt}
\caption{The curve of effective potential with $\ell$=0.60M, $b=3.9512M$ and $M=1$ in the regular Hayward black hole space-time. \label{v3} }
\end{figure}
\begin{figure}[H]
\centering
    \includegraphics[angle=0, width=0.39\textwidth]{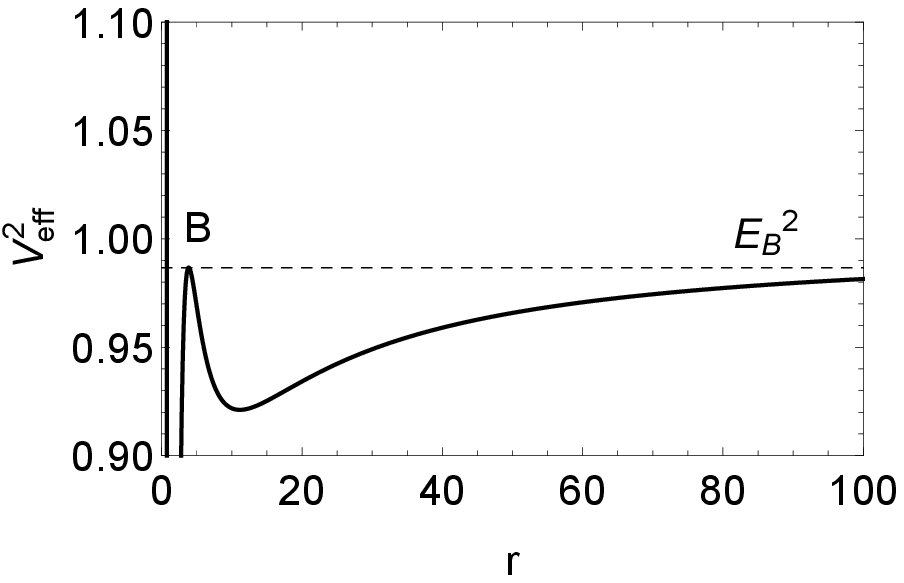}
     \includegraphics[angle=0, width=0.355\textwidth]{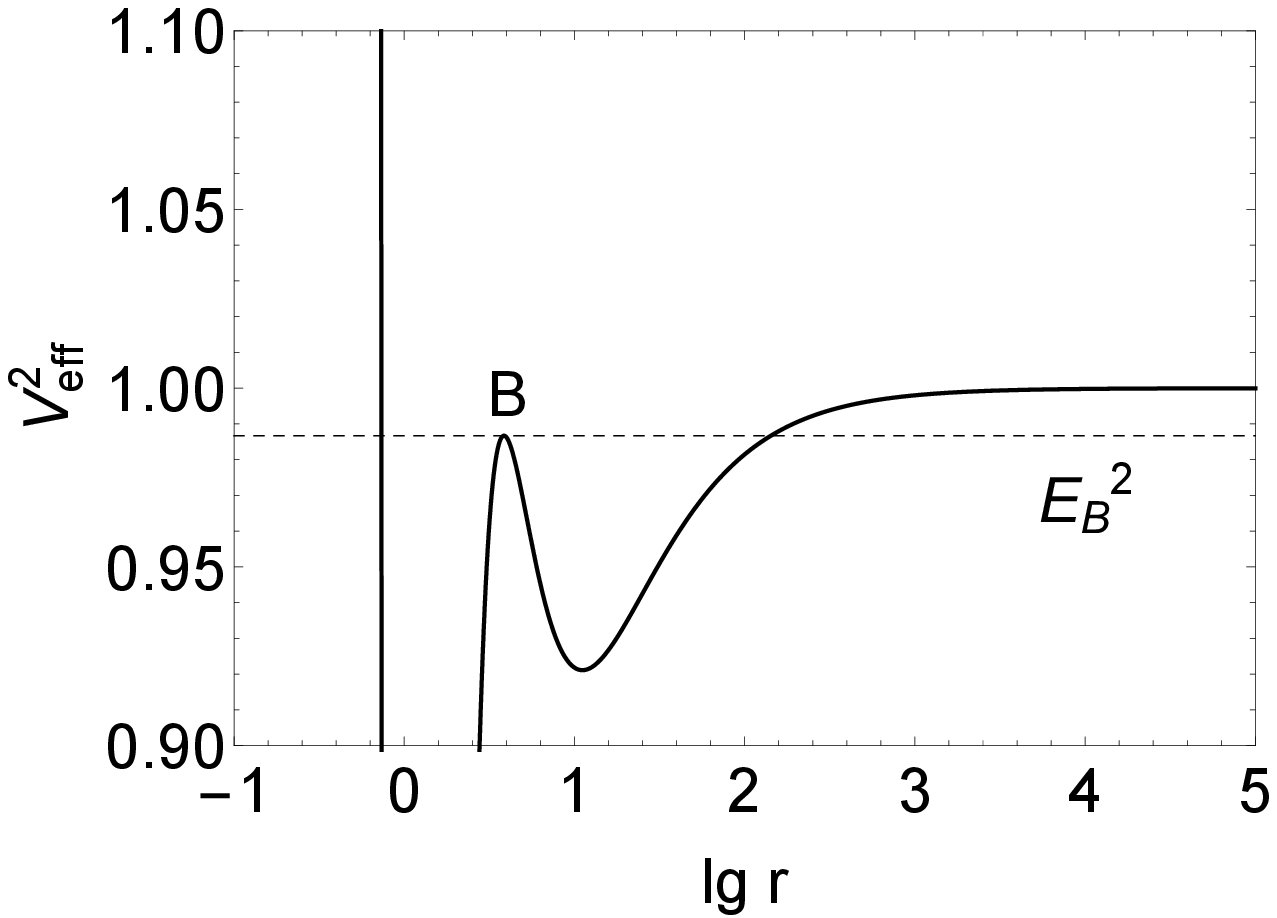}
     \vspace*{8pt}
\caption{The curve of effective potential with $\ell$=0.60M, $b$=3.90M and $M$=1 in the regular Hayward black hole space-time. \label{v4} }
\end{figure}
\begin{figure}[H]
\centering
    \includegraphics[angle=0, width=0.39\textwidth]{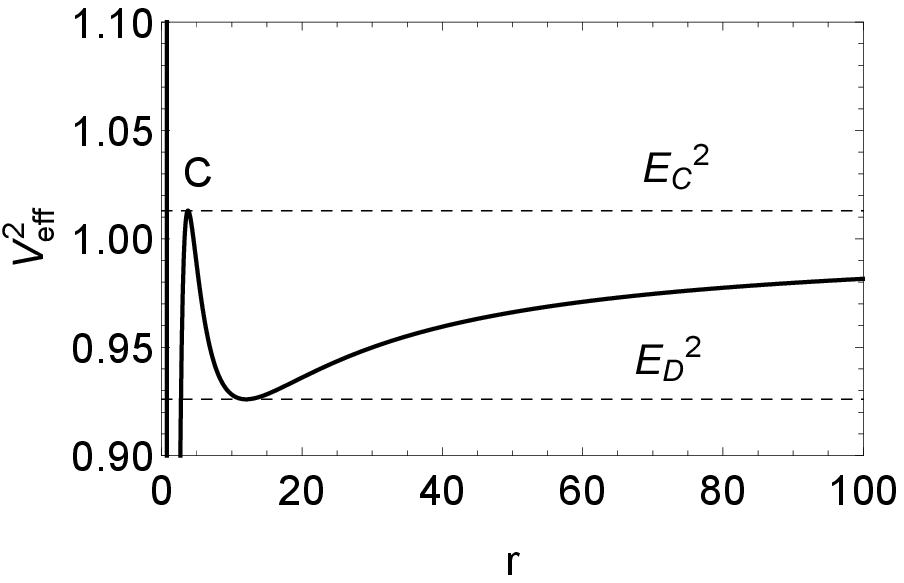}
     \includegraphics[angle=0, width=0.355\textwidth]{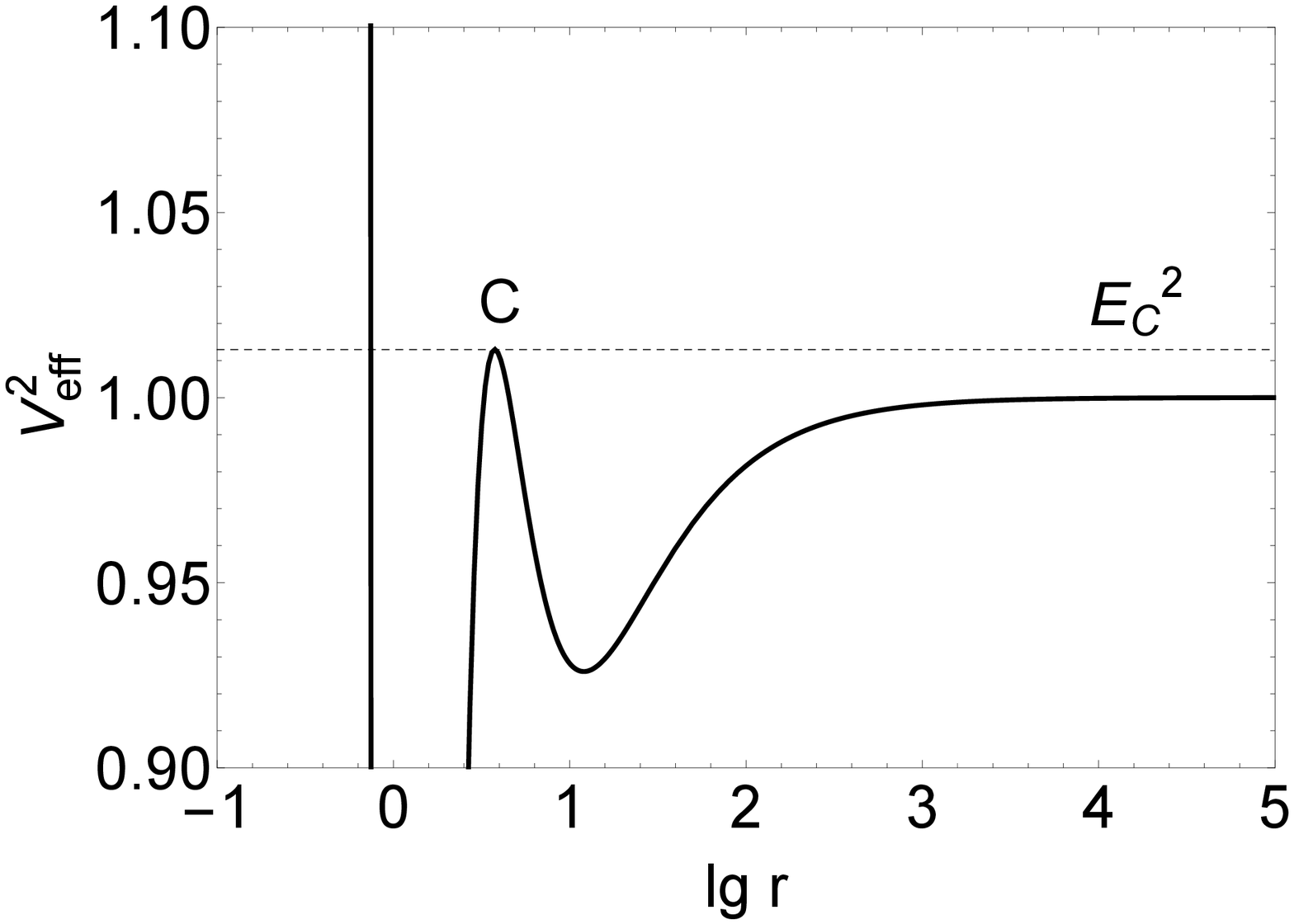}
     \vspace*{8pt}
\caption{The curve of effective potential with $\ell$=0.60M, $b$=4.00M and $M$=1 in the regular Hayward black hole space-time.\label{v5} }
\end{figure}

In Fig. \ref{v2}(a), three kinds of space-time with no horizon, one horizon and double horizons are revealed by analyzing the behaviours of effective potential obtained with different $\ell$. Fig. \ref{v2}(b) indicates that the stability of orbital motion is associated with the angular momentum $b$. No escape orbits exist when the values of $b$ are small enough. Based on Eq. \ref{e9}, a specific value $b$ = 3.9512$M$ is observed when $\ell = 0.60M$. We plot the curve of effective potential with $b$ =3.9512$M$ in Fig. \ref{v3}. By analyzing Eq. \ref{e9}, we find that the value of the effective potential approaches 1.00 as $r \rightarrow \infty $. For $b = 3.9512M$, the peak of effective potential equals to 1.00 (i.e. $V^{2}_{eff}(A) = 1.00$). When we choose $E^{2}$ $=$ $E^{2}_{A}$, the test particles move on unstable circular orbits, and the radius of circular orbits $r = r_{A}$. When the test particles are perturbed, they may escape from the circular orbits and then plunge into or fly away from the black hole. We also plot the curves of effective potential with $b$ = 3.90$M$ and $b$ = 4.00$M$ which are shown in Fig. \ref{v4} and \ref{v5}.

When $b$ is $3.90M$, $V^{2}_{eff}(B)$ is smaller than 1.00. If the test particles are thrown in black hole direction, they can only fall into the black hole with $E^{2}$ $>$ $E^{2}_{B}$. In this case, there are no escape orbits.
If $b$ is greater than $3.9512M$, $V^{2}_{eff}(C)$ is more than 1.00. For $E^{2} = E^{2}_{C}$, unstable circular orbits exist. When the test particles are perturbed, there are two different results. Under $r < r_{C}$, they will plunge into the black hole finally, and with $r > r_{C}$, they will move from $r_{C}$ to infinity. Under $E^{2}_{D} < E^{2} < E^{2}_{C}$, there are three kinds of orbits, bound orbits, escape orbits and absorbing orbits. By combining the Fig. \ref{v3}, \ref{v4} and \ref{v5}, it can be seen whether stable or unstable orbits exist is related to the parameter $b$ and that the peak of the effective potential is also associated with $b$.

\subsection{Time-like orbital motion with different values of $b$}
The corresponding equation of orbital motion becomes
\begin{eqnarray}
\dot{r}^{2}=E^{2}-(1-\frac{2Mr^{2}}{r^{3}+2Ml^{2}})(1+\frac{b^{2}}{r^{2}}).
\label{e10}
\end{eqnarray}
Letting $R=\frac{1}{r}$, a new expression can be derived from Eq. (\ref{e10})
 \begin{eqnarray}
 (\frac{dR}{d\phi})^{2}=\frac{E^{2}-1}{b^{2}}-R^{2}+\frac{2MR+2Mb^{2}R^{3}}{(1+2M\ell^{2}R^{3})b^{2}}.
\label{e11}
\end{eqnarray}
The second order motion equation can be obtained by differentiating Eq. (\ref{e11})
\begin{eqnarray}
 \frac{d^{2}R}{d\phi^{2}}+R=\frac{3MR^{2}b^{2}+3M}{b^{2}(1+2M\ell^{2}R^{3})^{2}}-\frac{2M}{(1+2M\ell^{2}R^{3})b^{2}}.
\label{e12}
\end{eqnarray}
By solving Eq. (\ref{e11}) and Eq. (\ref{e12}) numerically, all possible orbits of massive particles can be found near $b=3.9512M$ in the regular Hayward black hole space-time. We also examine the influence of parameters ($E^{2}$ and $b$) on the time-like geodesics in the regular Hayward black hole space-time.

\begin{figure}[H]
\centering
      \includegraphics[angle=0, width=0.34\textwidth]{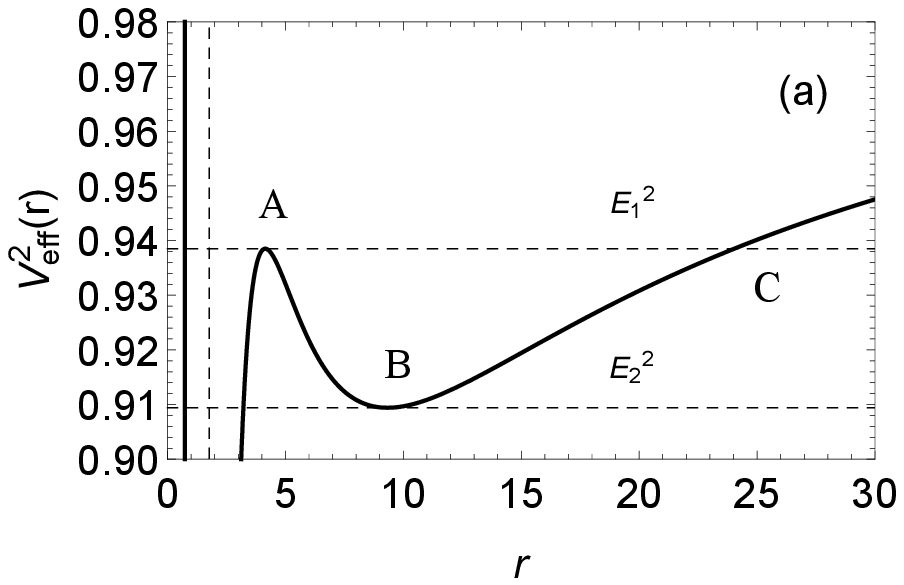}
     \ \ \ \ \ \includegraphics[angle=0, width=0.34\textwidth]{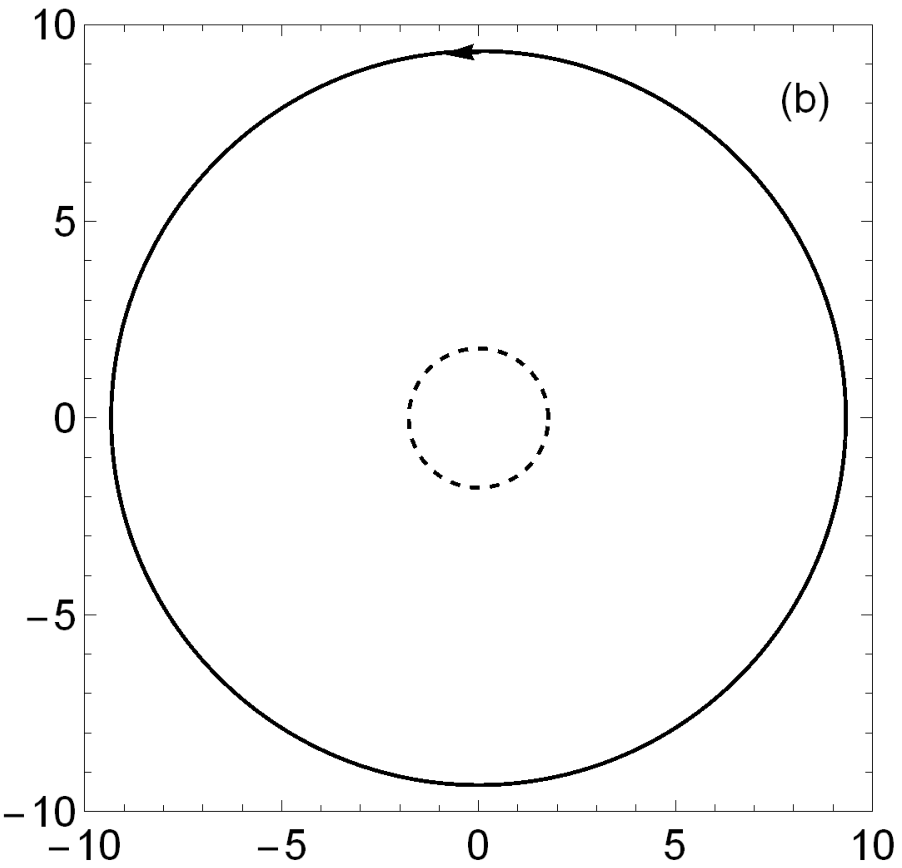}
     \\ \ \includegraphics[angle=0, width=0.32\textwidth]{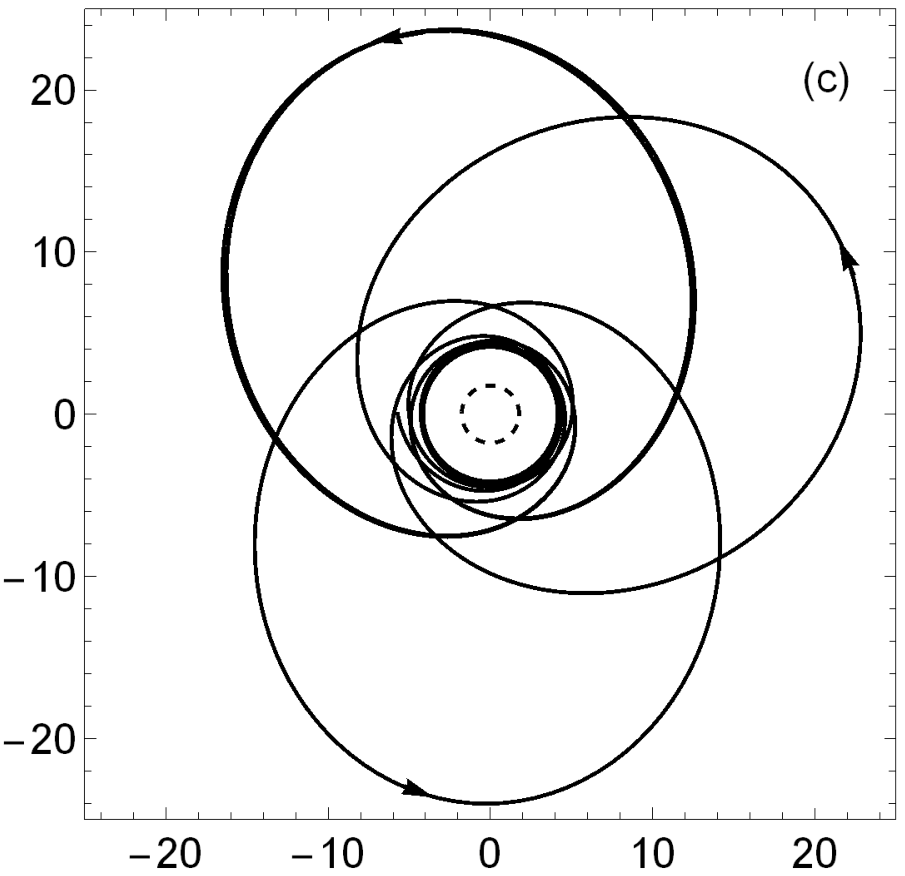}
    \ \ \ \ \ \ \includegraphics[angle=0, width=0.32\textwidth]{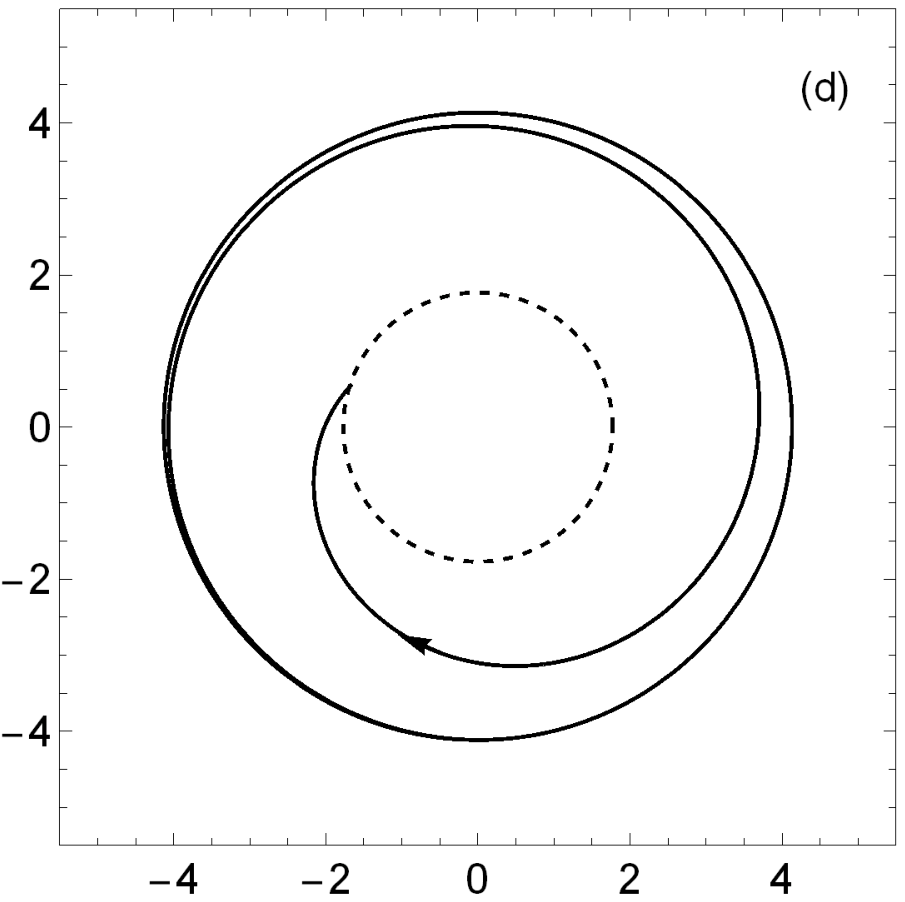}
      \\ \ \  \ \includegraphics[angle=0, width=0.31\textwidth]{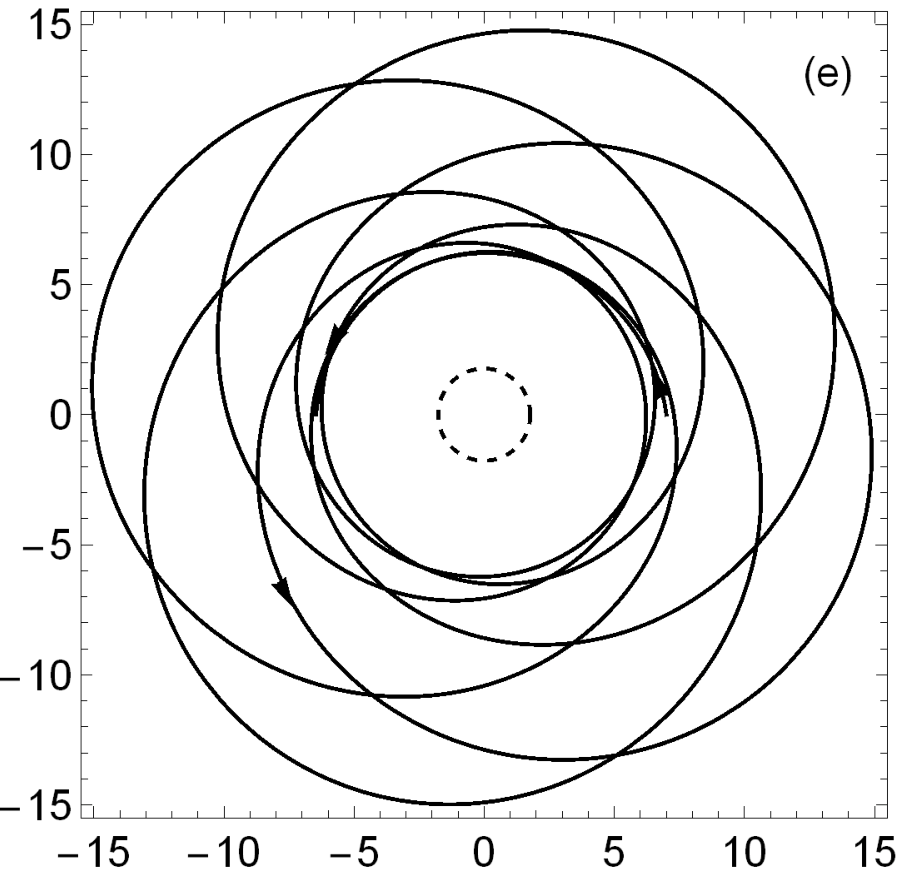}
    \  \ \ \ \ \ \ \includegraphics[angle=0, width=0.33\textwidth]{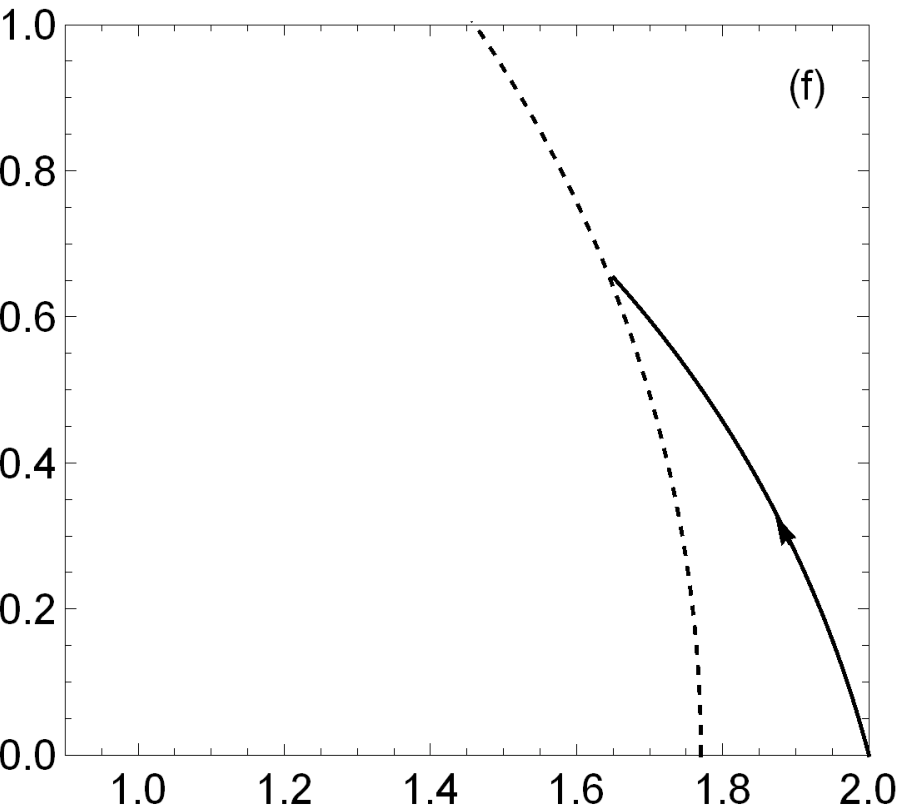}
      \\ \ \ \includegraphics[angle=0, width=0.316\textwidth]{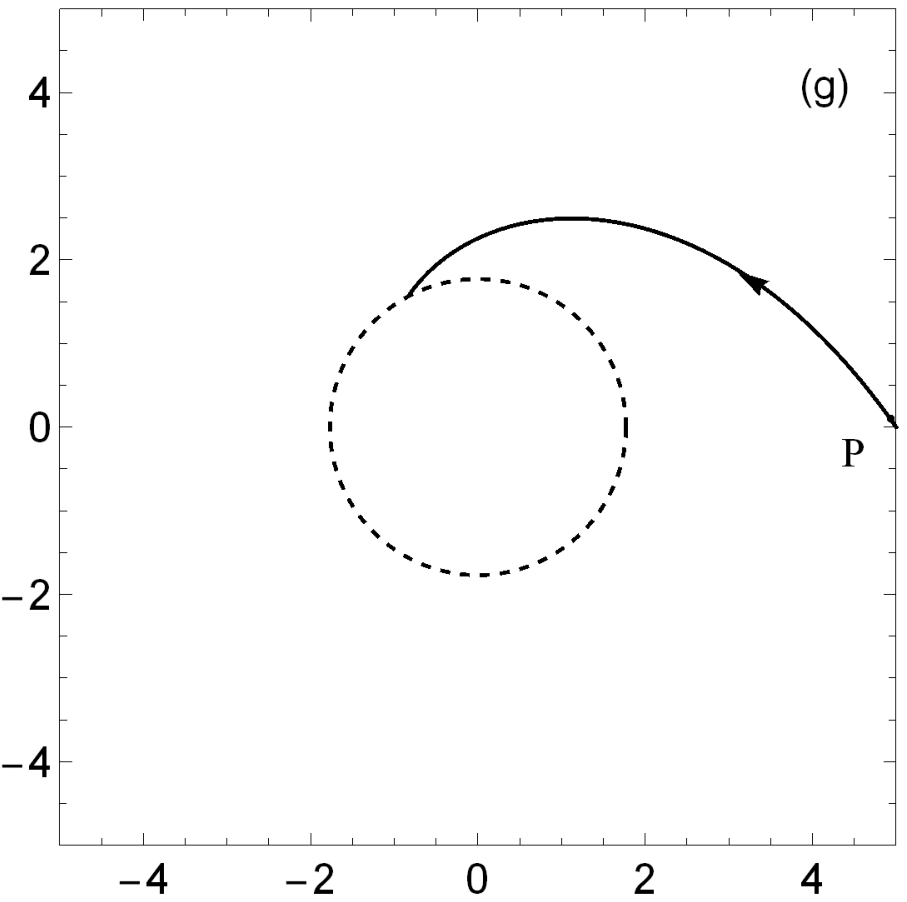}
     \ \ \ \ \ \ \includegraphics[angle=0, width=0.33\textwidth]{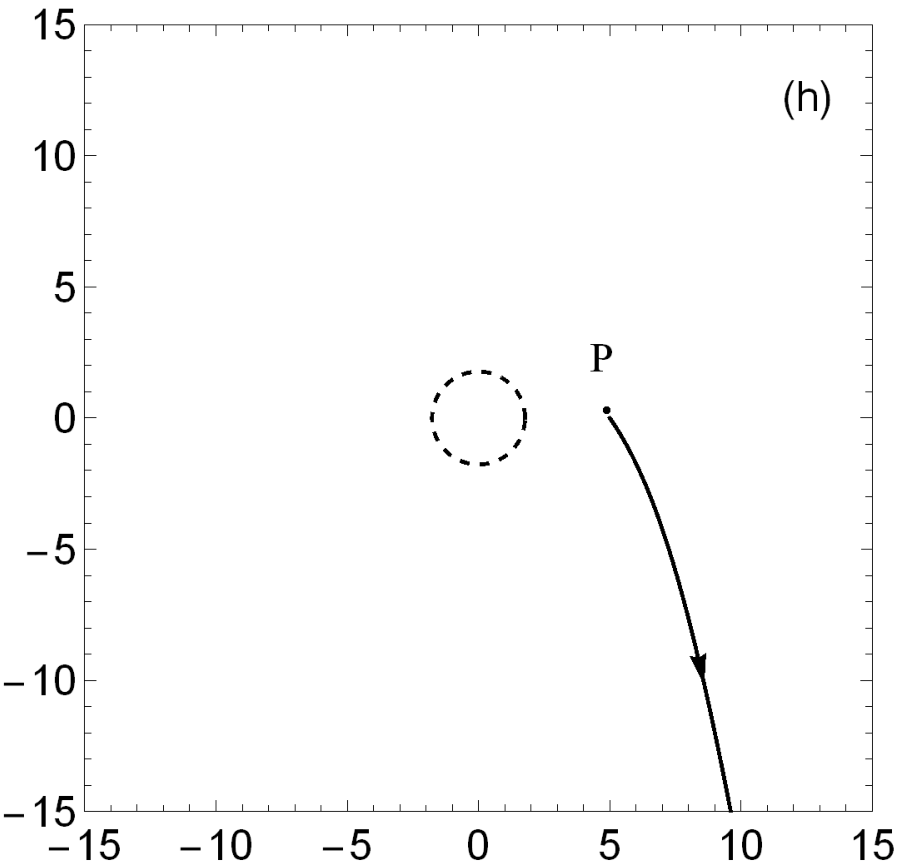}
     \vspace*{8pt}
\caption{All possible time-like orbits with $\ell=0.60M$, $b$=3.70$M$ and $M=1$ in the regular Hayward black hole space-time. \label{v6} }
\end{figure}

In Fig. \ref{v6}, all possible orbits are demonstrated for $b = 3.70M$. We can find that: (1) When the energy level $E^{2}$ = $E^{2}_{2}$, the orbits are stable circular orbits ($r = r_{B}$). It is shown in Fig. \ref{v6} (b). (2) When the energy level $E^{2}$ = $E^{2}_{1}$, the test particles can move on unstable circular orbits. If $r > r_{A}$, any perturbation would cause them to go off the circular orbit. Then they will move on bound orbits between $r_{A}$ and $r_{C}$. For $r < r_{A}$, any perturbation would cause the particles to break away from the circular orbit and eventually fall into the black hole. These two cases are illustrated in Fig. \ref{v6} (c) and (d). (3)When $E^{2}_{2} < E^{2} < E^{2}_{1}$, there exist two kinds of orbits, bound orbits and absorbing orbits which are shown in Fig. \ref{v6} (e) and (f). In addition, another two kinds of orbits are discovered. (4)When $E^{2}$ $>$ $E^{2}_{1}$, if we throw the test particles in direction of black hole from point P, the test particles will fall into the black hole due to not enough angular momentum $b$, as shown in Fig. \ref{v6} (g). If we throw the test particles in the opposite direction of black hole from point P, they will keep away from the black hole, as shown in Fig. \ref{v6} (h).From the above discussion, it can be concluded that $b = 3.70M$ is too small for the existence of escape orbits. Then we choose $b = 5.00M$ to plot all possible orbits.

\begin{figure}[H]
\centering
      \includegraphics[angle=0, width=0.34\textwidth]{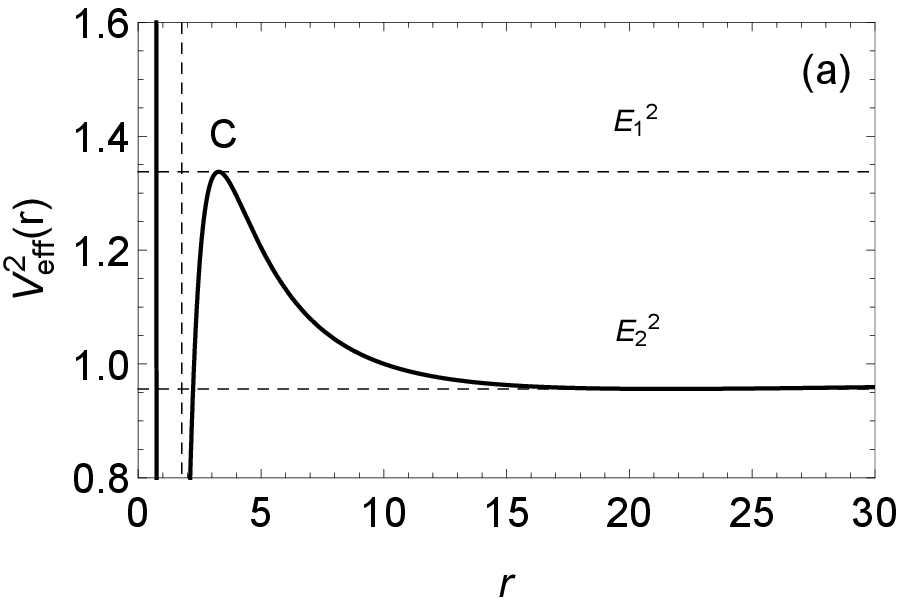}
     \ \ \ \  \includegraphics[angle=0, width=0.34\textwidth]{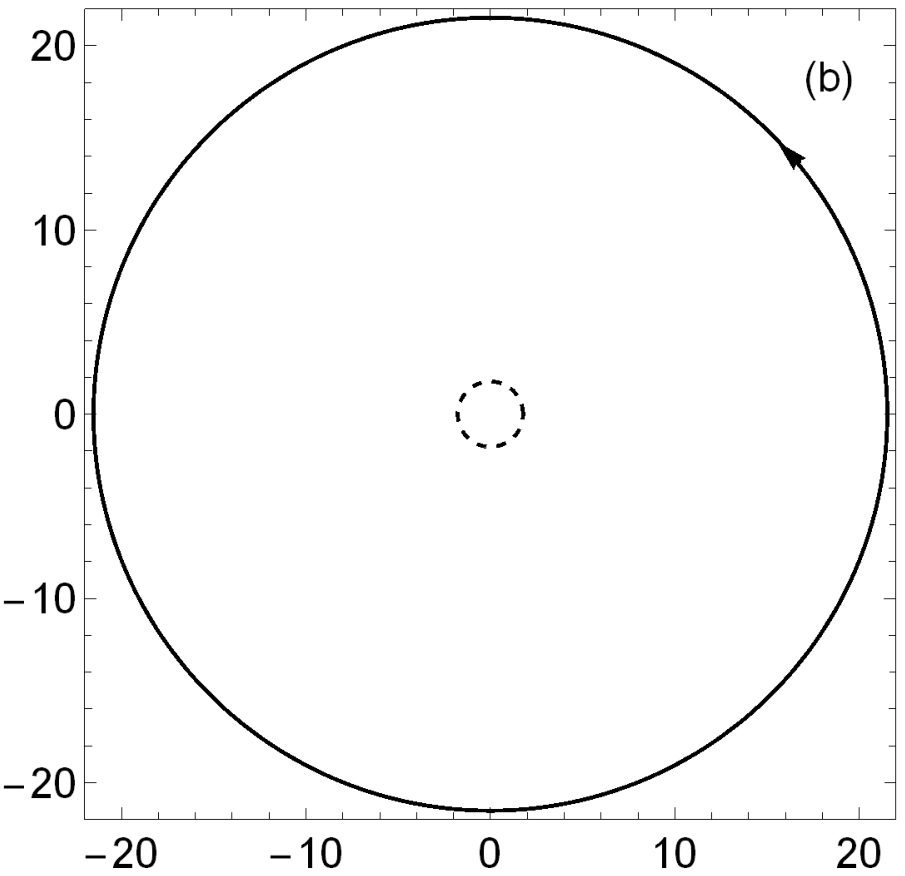}
     \\ \includegraphics[angle=0, width=0.32\textwidth]{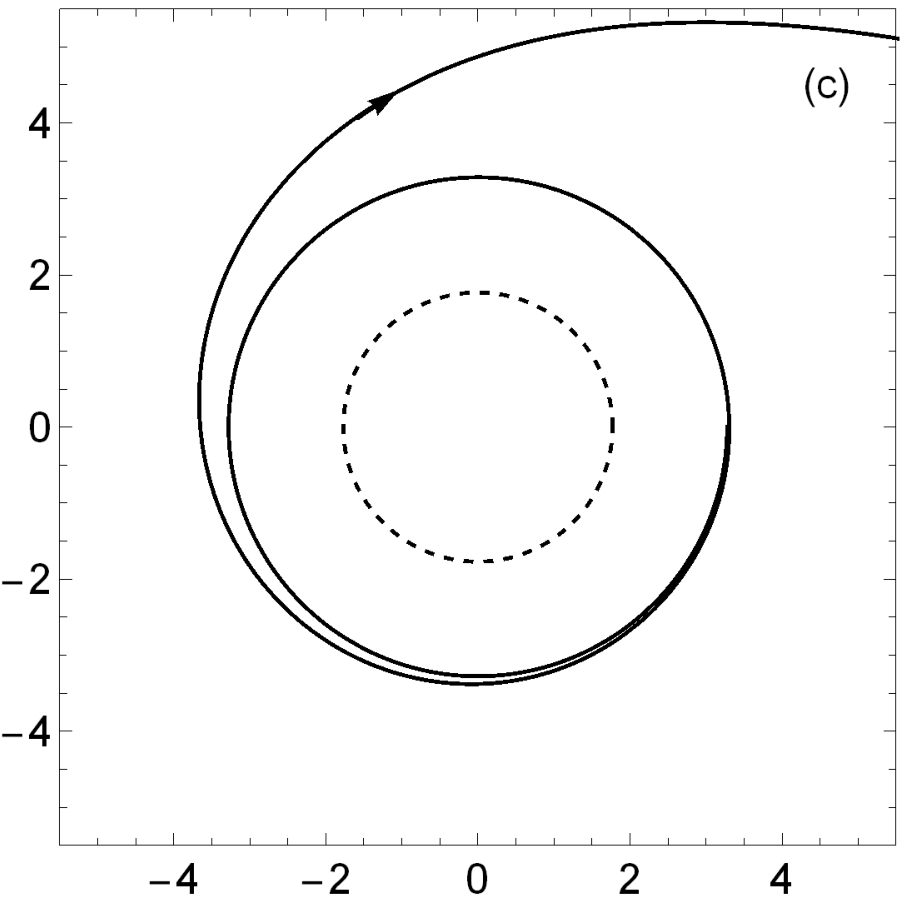}
     \ \ \ \ \ \  \includegraphics[angle=0, width=0.33\textwidth]{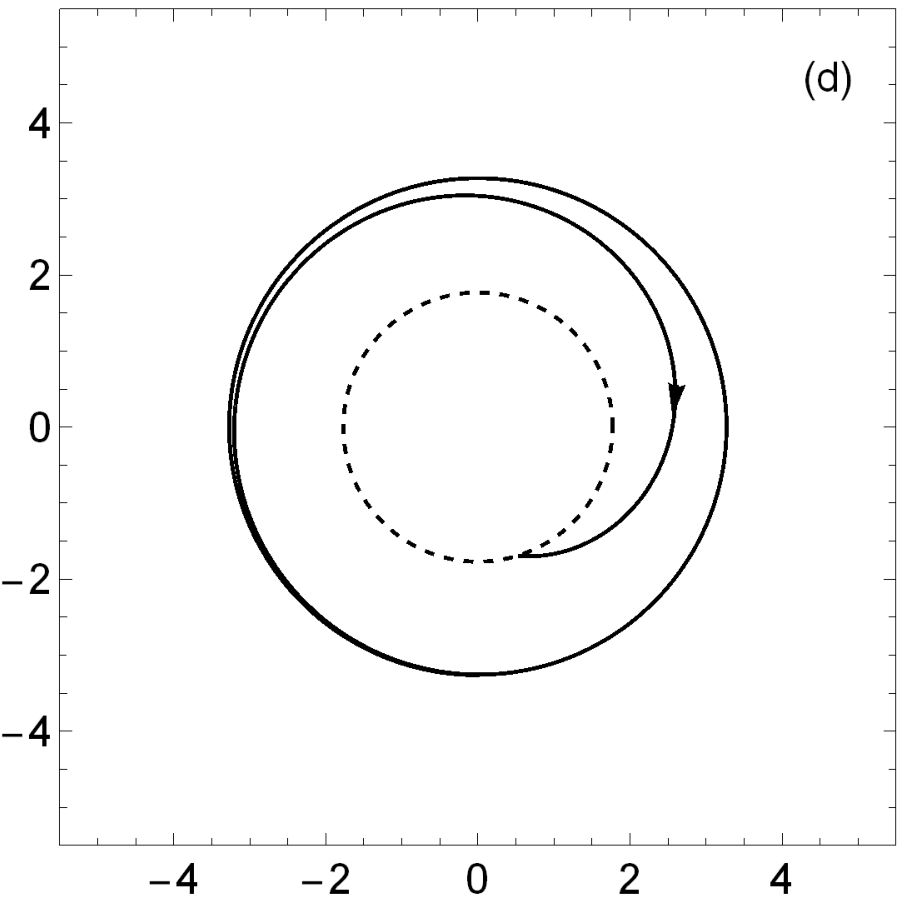}
      \\ \includegraphics[angle=0, width=0.31\textwidth]{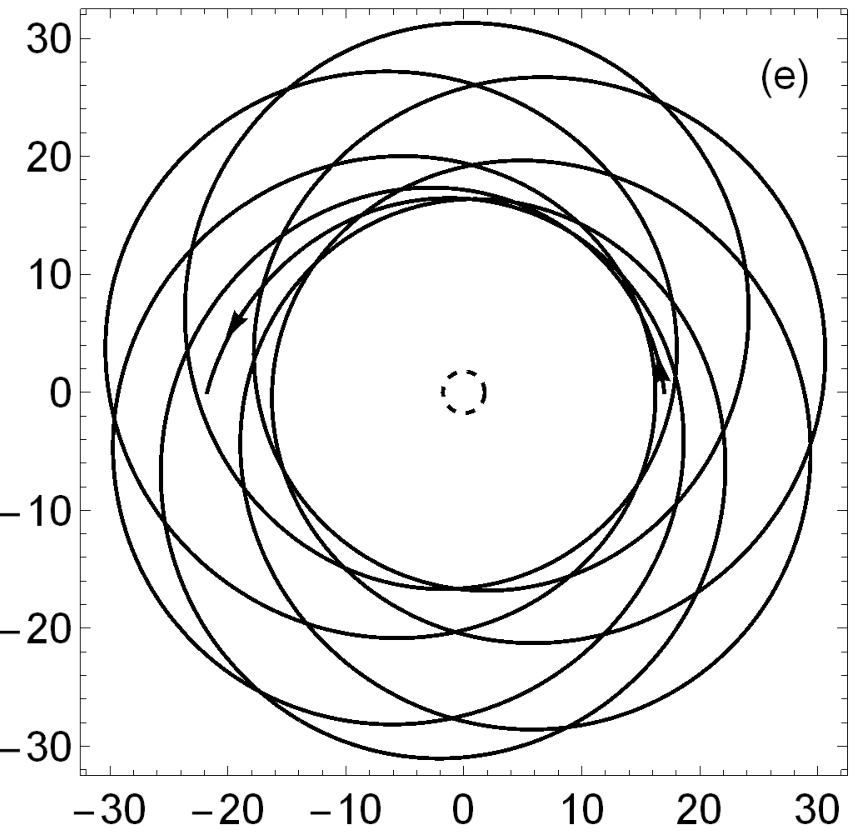}
     \ \ \ \ \ \ \includegraphics[angle=0, width=0.327\textwidth]{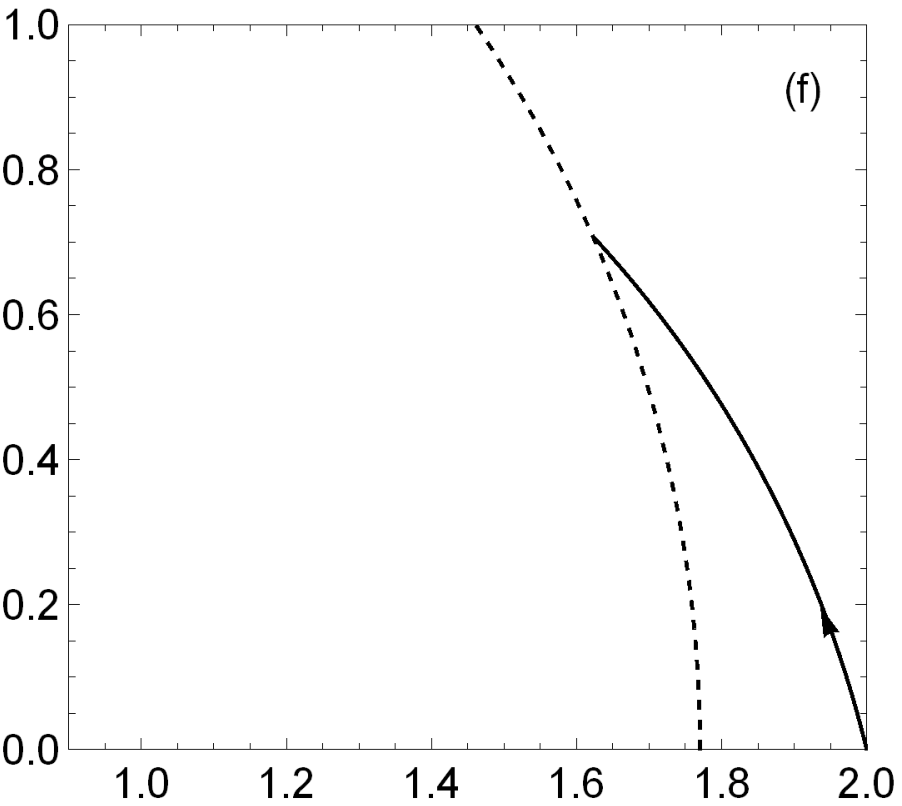}
     \\ \includegraphics[angle=0, width=0.33\textwidth]{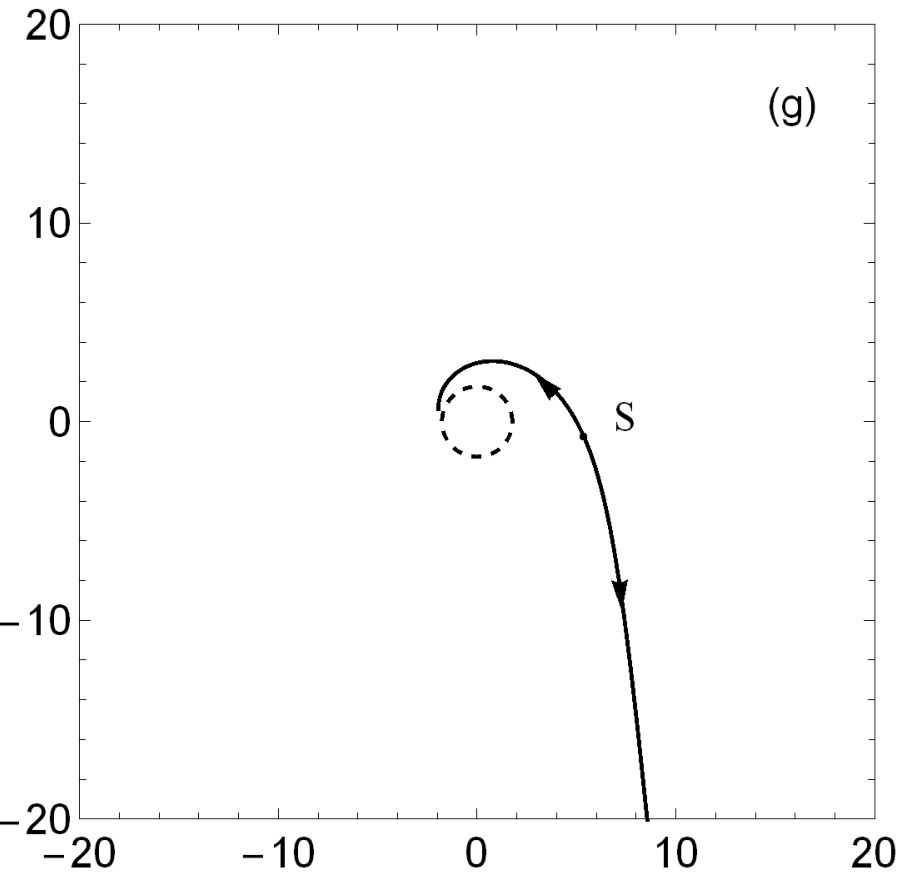}
     \ \ \ \  \includegraphics[angle=0, width=0.33\textwidth]{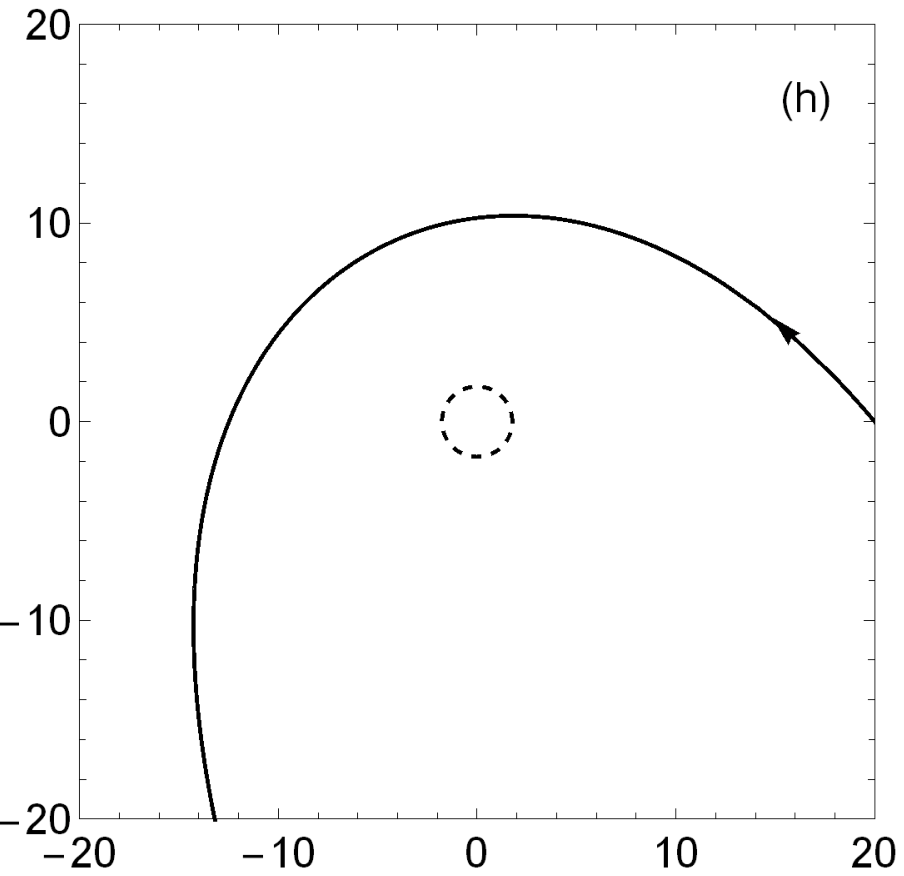}
     \vspace*{8pt}
\caption{All possible time-like orbits with $\ell=0.60M$, $b$=5.00$M$ and $M=1$ in the regular Hayward black hole space-time. \label{v7} }
\end{figure}

Comparing Fig. \ref{v6} with Fig. \ref{v7}, we can find some differences with different values of angular momentum ($b = 3.70M$ and $b = 5.00M$) as follows:
\begin{enumerate}
  \item As $b$ increases, the radii of stable circular orbits become larger ($r_{B} < r_{D}$) but the radii of unstable circular orbits become smaller ($r_{C} < r_{A}$).
  \item The range of the bound orbits and absorbed orbits increases with the increase of $b$ ($\Delta E^{2}_{1}< \Delta E^{2}_{2}$).
  \item The orbital types under $b = 3.7M$ do not include the escape orbits. The escape orbits are shown in Fig. \ref{v7} (h)
  \item The peak of effective potential $V^{2}_{eff}$ becomes higher as $b$ increases ($ E^{2}_{A}<  E^{2}_{C}$).
  \item The velocity of bound orbits increases with the increase of $b$. From Fig. \ref{v6}$(b=3.70M)$ and Fig. \ref{v7}$(b=5.00M)$, we also find that their procession direction are different. The precession direction of Fig. \ref{v6} is clockwise while that of Fig. \ref{v7} is counterclockwise.
  \item The unstable circular orbits in Fig. \ref{v6} (c) include bound orbits shifting clockwise but the unstable orbits of Fig. \ref{v7} (c) include escapes orbits from $r_{C}$ to infinity.
  \end{enumerate}
\subsection{The influence of energy level $E^{2}$ on the orbital motion}
\begin{figure}[H]
  \centering
    \includegraphics[angle=0, width=0.25\textwidth]{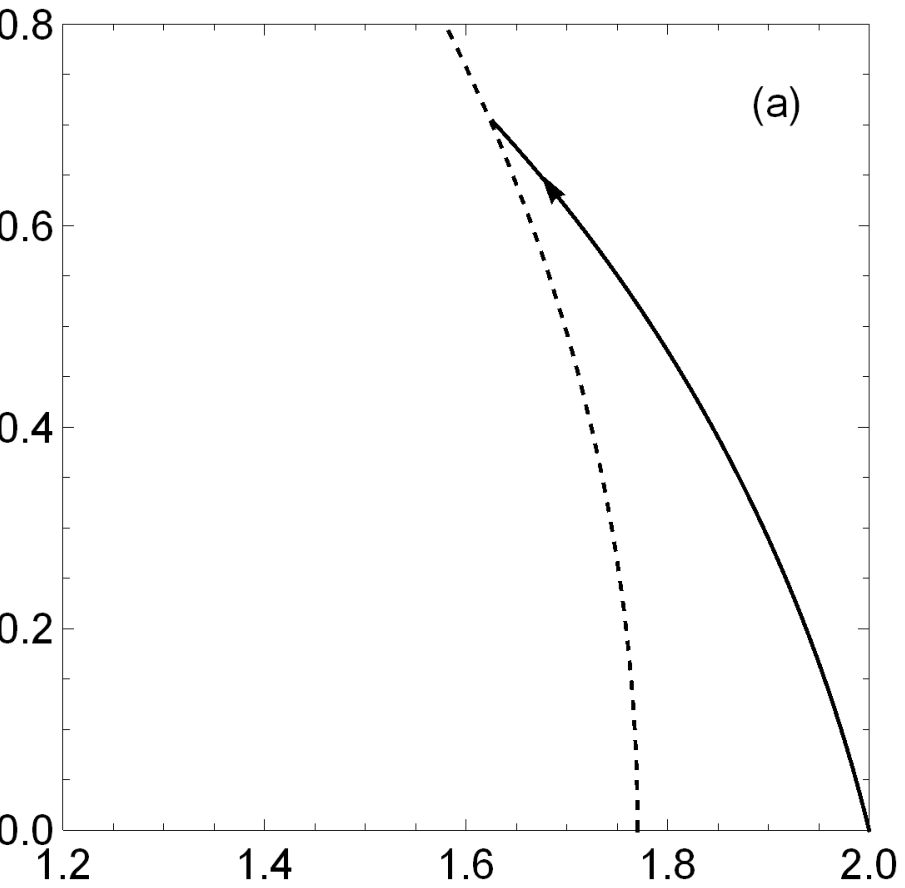}
    \ \ \ \ \ \includegraphics[angle=0, width=0.25\textwidth]{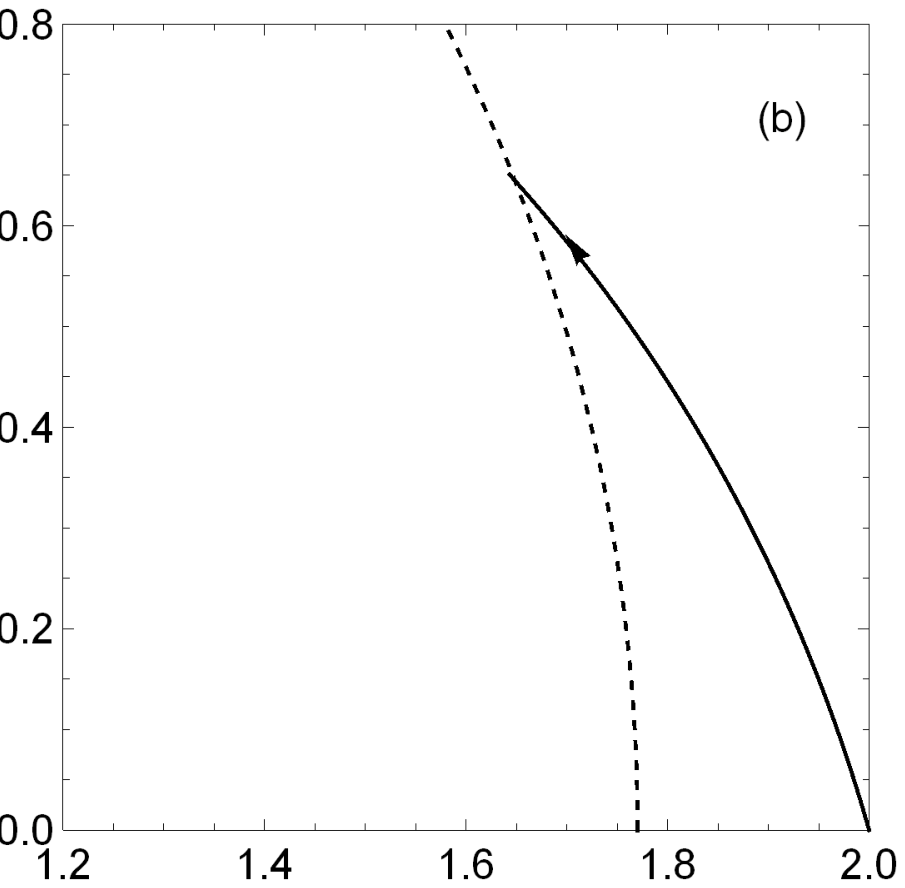}\\
     \ \includegraphics[angle=0, width=0.25\textwidth]{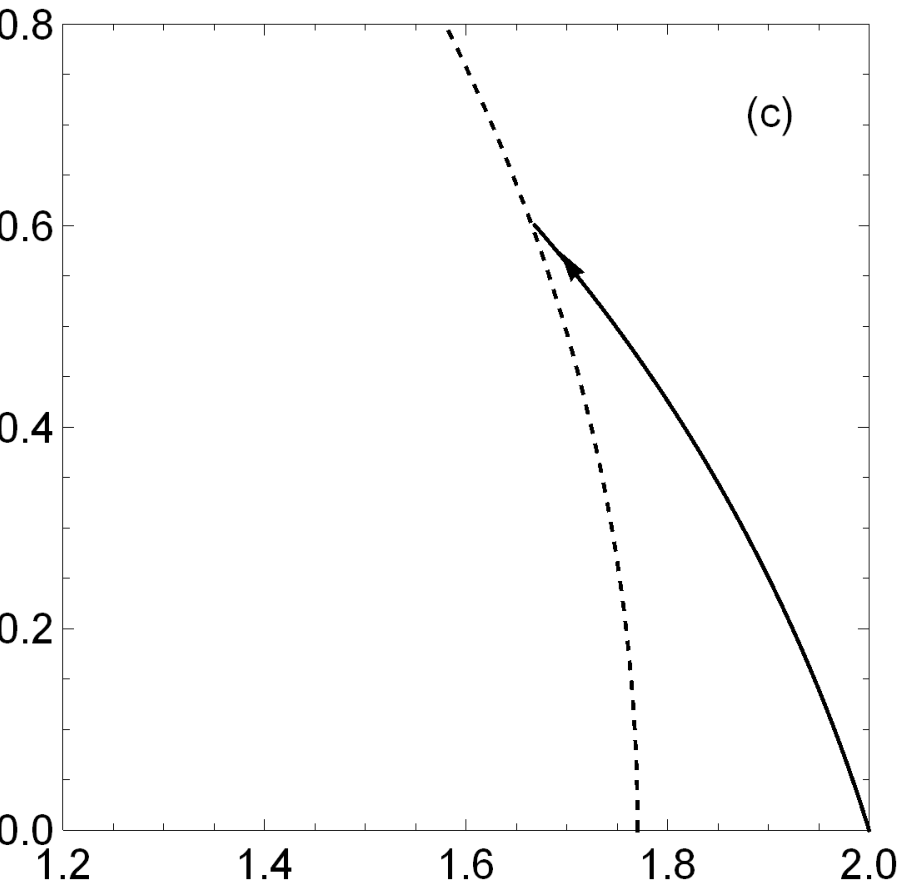}
    \ \ \ \ \ \includegraphics[angle=0, width=0.25\textwidth]{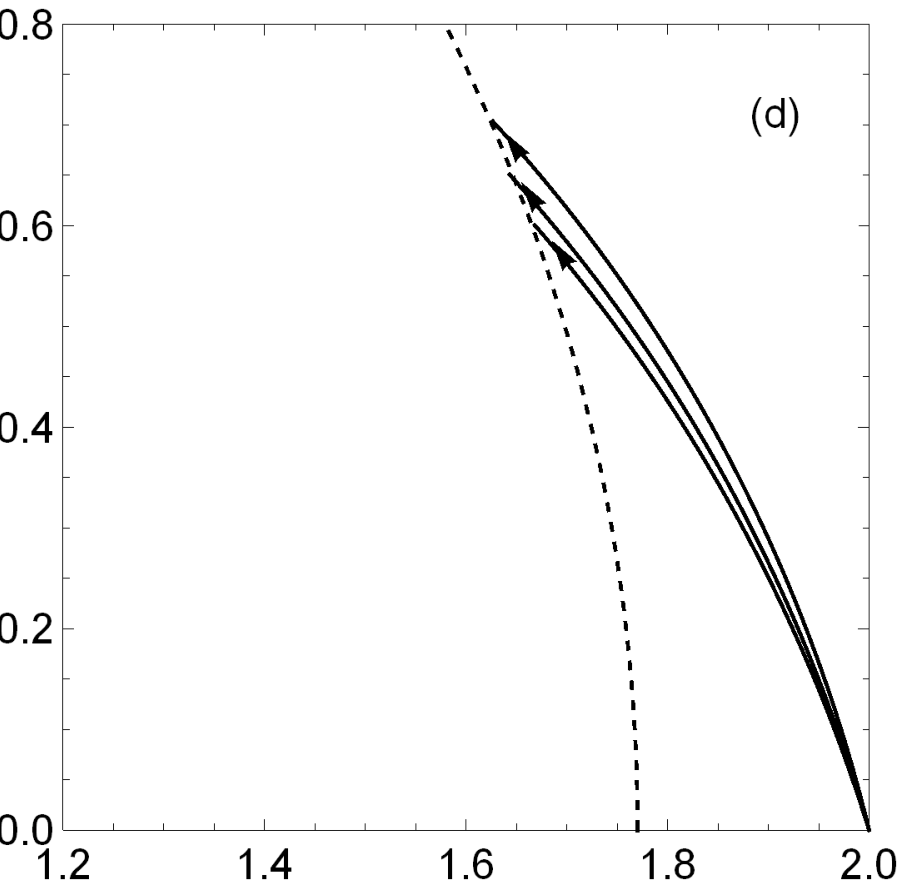}
     \vspace*{8pt}
\caption{The behavior of absorbing orbits with different values of energy level $E^{2}_{a}$=0.97, $E^{2}_{b}$=1.10 and $E^{2}_{c}$=1.20 for fixed $b=5.00M$, $\ell=0.60M$ and $M=1$. \label{v10} }
\end{figure}

When the angular momentum is determined, it can be known whether there exist stable or unstable orbits by analyzing the behavior of the corresponding effective potential. The type of orbital motion is determined by the energy level $E^{2}$. Here, detailed analyses are made to discuss the influence of energy level on the same orbital type, as illustrated in  Fig. \ref{v8} , \ref{v9} and \ref{v10}.
\subsubsection{The bound geodesics}
\begin{figure}[H]
\centering
    \includegraphics[angle=0, width=0.3\textwidth]{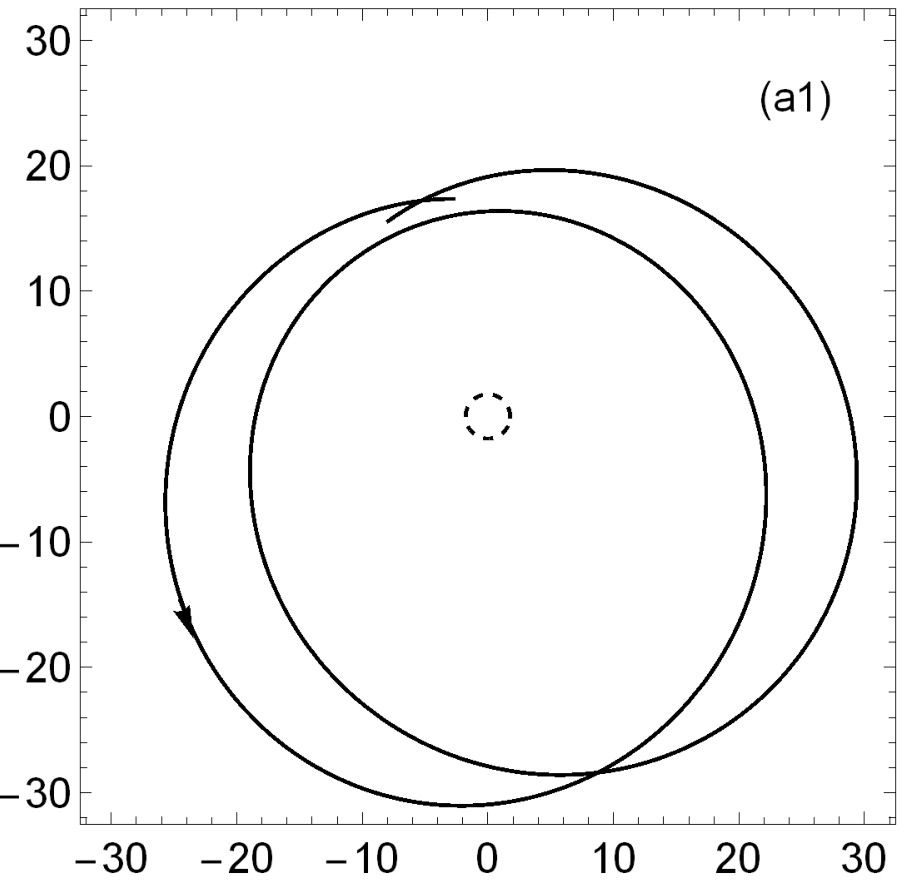}
    \ \ \ \ \ \includegraphics[angle=0, width=0.3\textwidth]{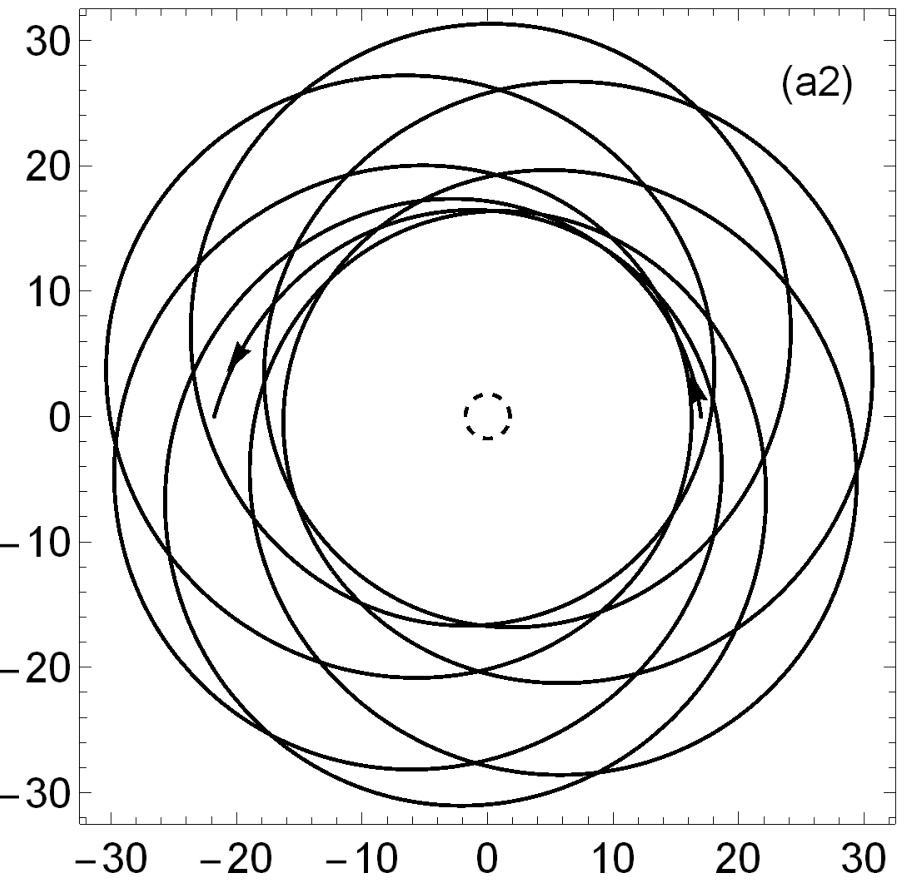}\\
    \ \ \includegraphics[angle=0, width=0.31\textwidth]{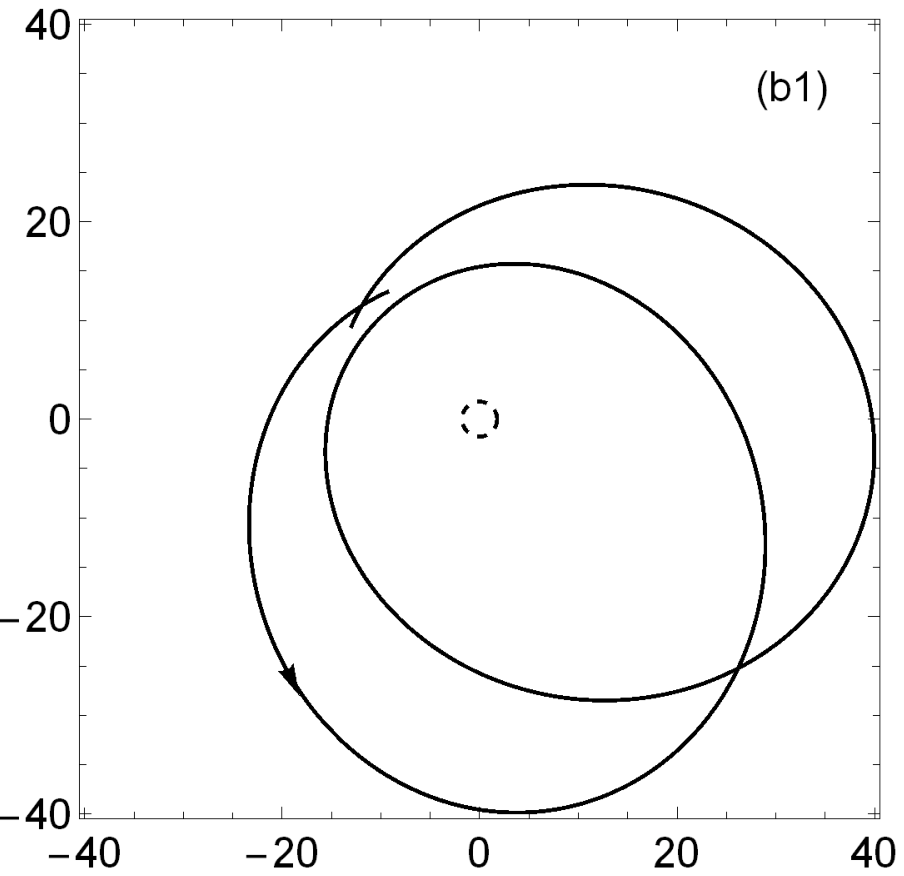}
    \ \ \ \ \includegraphics[angle=0, width=0.31\textwidth]{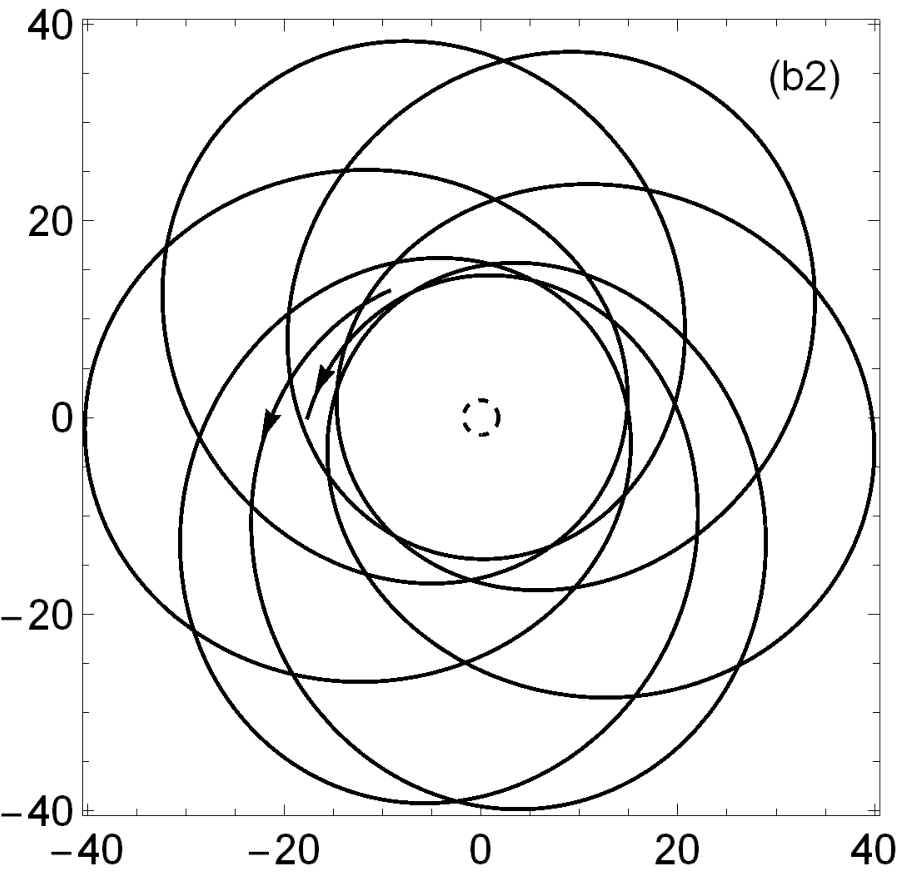}
     \vspace*{8pt}
\caption{The behavior of bound orbits with different values of energy level $E^{2}_{a}$=0.96 and $E^{2}_{b}$=0.965 for fixed $b=5.00M$, $\ell=0.60M$ and $M=1$. \label{v8} }
\end{figure}

In Fig. \ref{v8}, we plot the bound orbits with different values of energy level and find that it has some influences on the radius of orbital motion and the precession velocity. Figs. \ref{v8} a1, a2 have higher energy level, bigger radius and higher precession velocity than Fig. \ref{v8} b1, b2.
\subsubsection{The escape geodesics}
\begin{figure}[H]
  \centering
    \includegraphics[angle=0, width=0.3\textwidth]{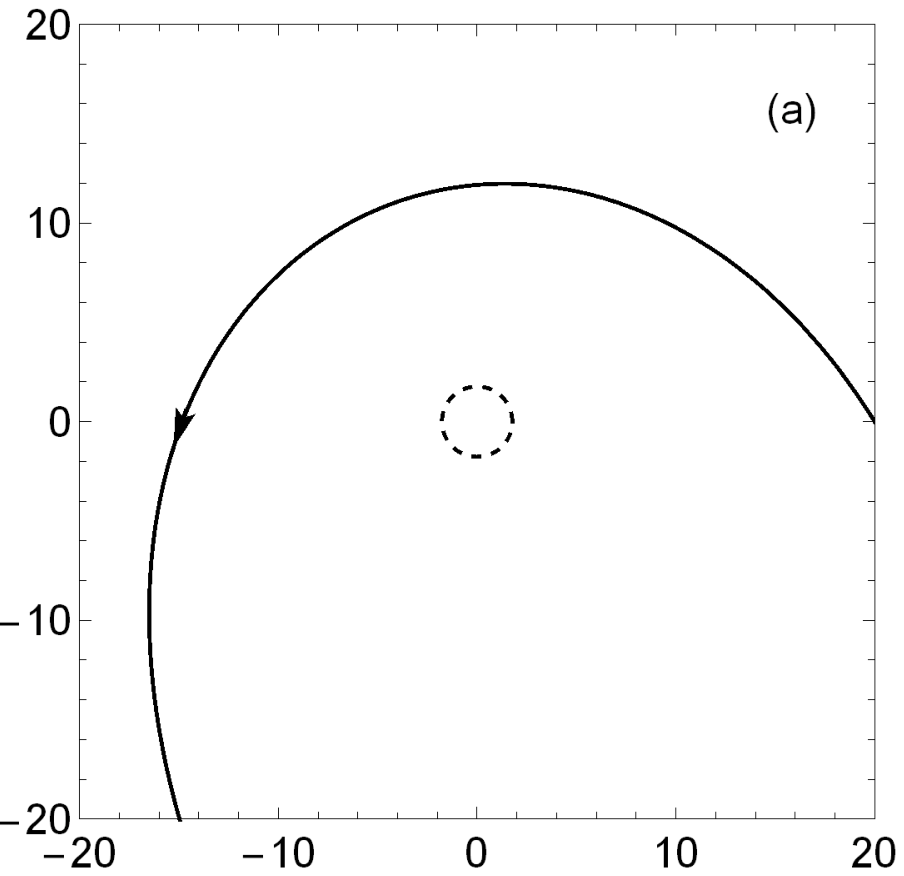}
    \ \ \ \ \ \includegraphics[angle=0, width=0.3\textwidth]{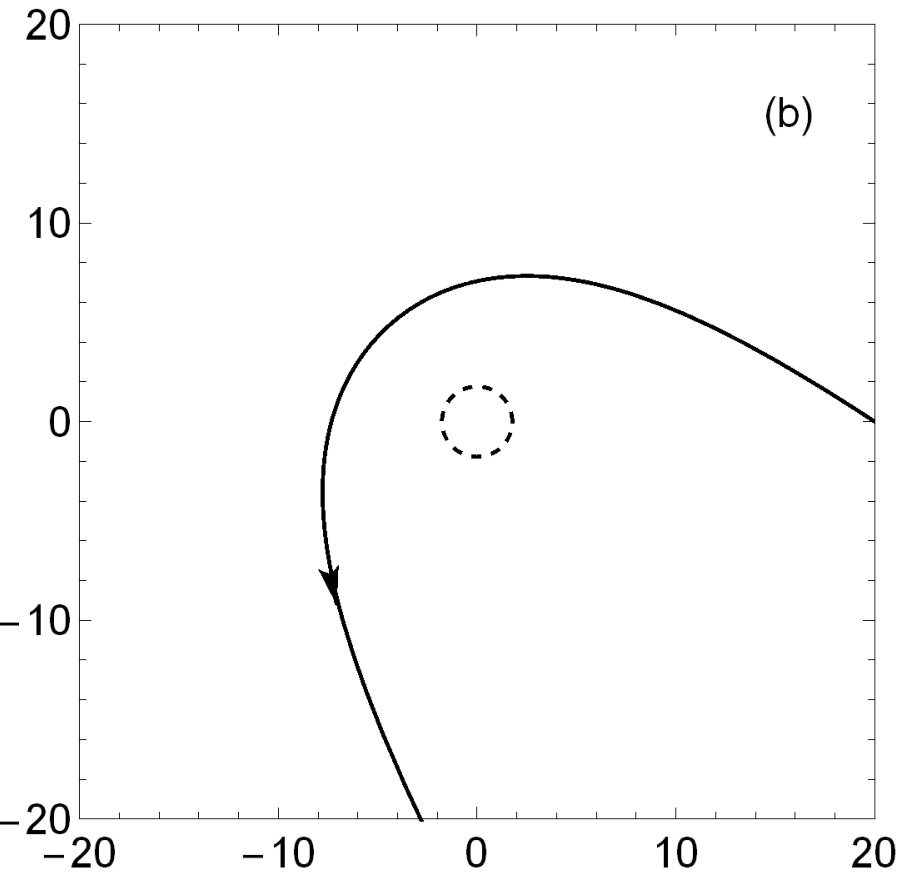}\\
     \ \includegraphics[angle=0, width=0.3\textwidth]{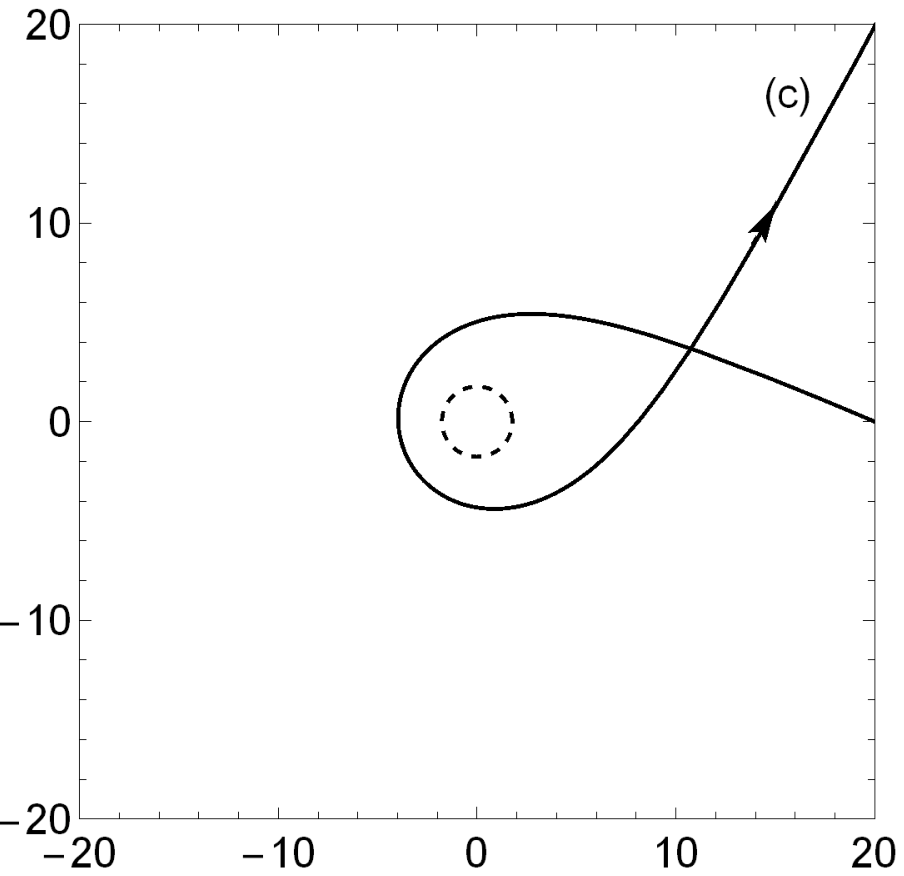}
     \ \ \ \ \ \includegraphics[angle=0, width=0.3\textwidth]{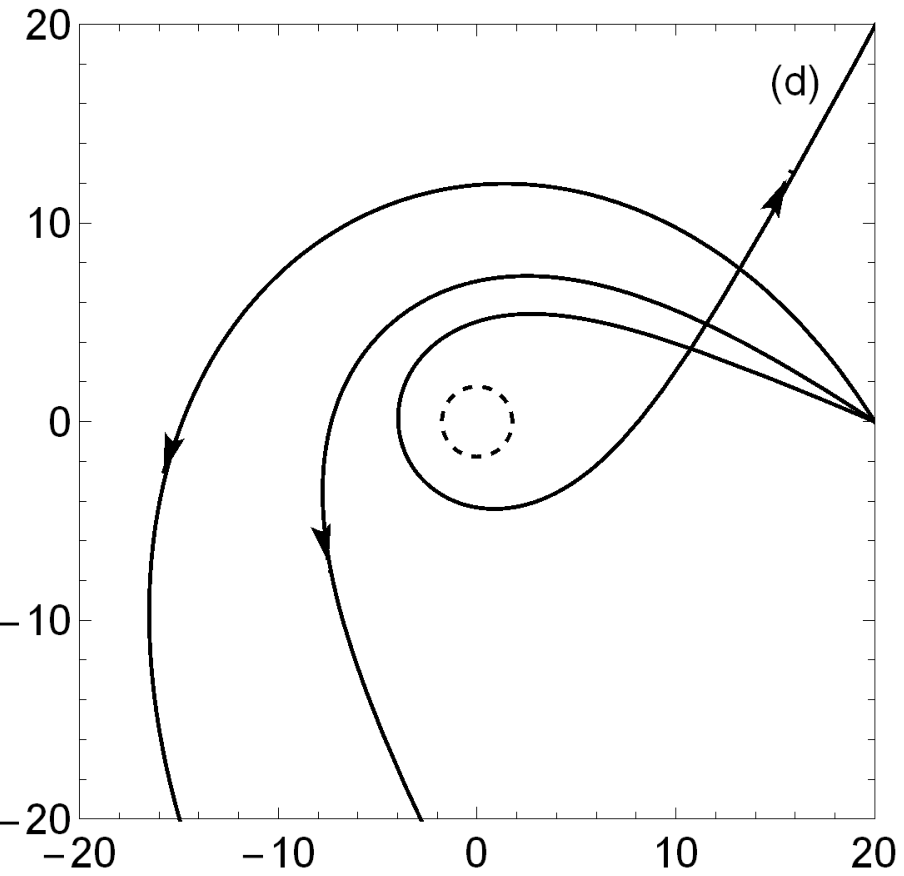}
     \vspace*{8pt}
\caption{The behavior of escape orbits with different values of energy level $E^{2}_{a}$=0.98, $E^{2}_{b}$=1.10 and $E^{2}_{c}$=1.20 for fixed $b=5.00M$, $\ell=0.60M$ and $M=1$. \label{v9} }
\end{figure}

In Fig. \ref{v9}, if we throw test particles in the direction of black hole, their orbits will undergo a certain deflection. With the increase of the energy level, the curvature of the orbits increases and they eventually intersect themselves.

\subsubsection{The absorbing geodesics}
If we throw the test particles near the horizon in the direction black hole, they will always plunge into the black hole, and those with higher energy will fall into the black hole earlier.
\section{The structure of null geodesics}
For the null geodesics $h=0$, the corresponding equation of effective potential is expressed as follows
\begin{eqnarray}
 V_{eff}^{2}=(1-\frac{2Mr^{2}}{r^{3}+2Ml^{2}})\frac{b^{2}}{r^{2}}.
\label{e13}
\end{eqnarray}
\begin{figure}[H]
\centering
    \includegraphics[angle=0, width=0.4\textwidth]{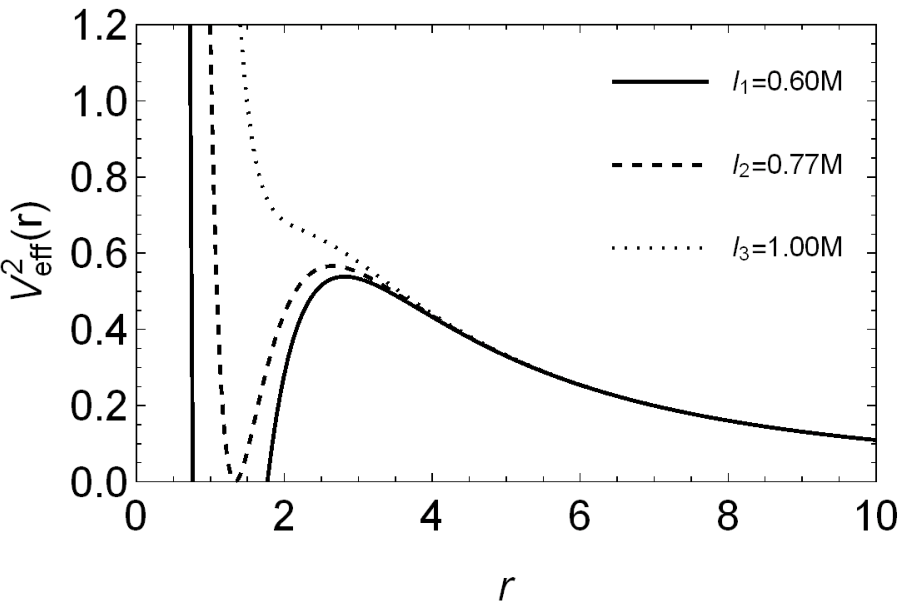}
    \ \ \ \  \includegraphics[angle=0, width=0.4\textwidth]{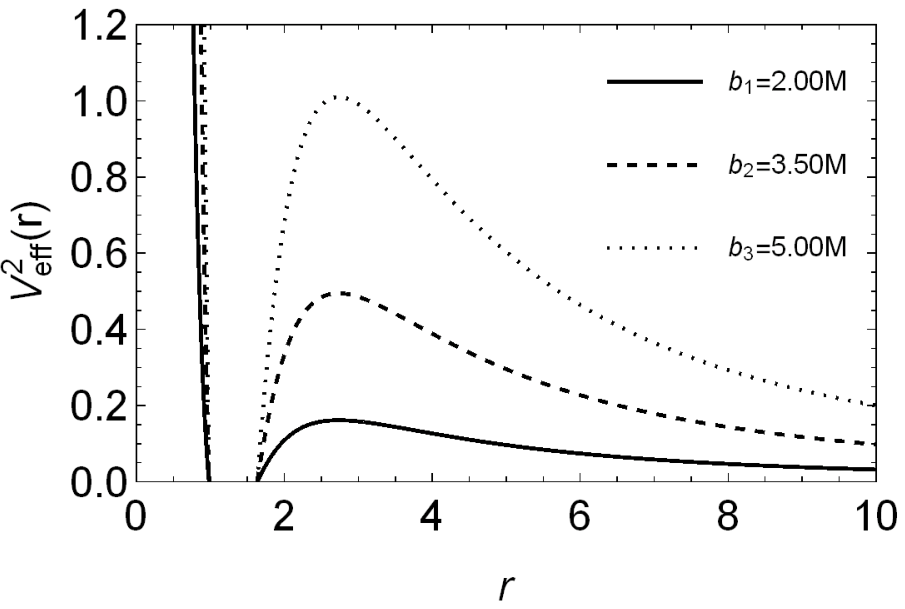}
     \vspace*{8pt}
\caption{The curves of effective potential with different values of parameters for fixed $b=3.70M$ (left) and fixed $\ell=0.80M$ (right). \label{v11} }
\end{figure}
The corresponding equation of orbital motion is
\begin{eqnarray}
\dot{r}^{2}=E^{2}-\frac{b^{2}}{r^{2}}(1-\frac{2Mr^{2}}{r^{3}+2Ml^{2}}).
\label{e14}
\end{eqnarray}
Let $R=\frac{1}{r}$, the equation of orbital motion becomes
\begin{eqnarray}
 (\frac{dR}{d\phi})^{2}=\frac{E^{2}}{b^{2}}-R^{2}+\frac{2MR^{3}}{1+2M\ell^{2}R^{3}}.
\label{e15}
\end{eqnarray}
Differentiating Eq. (\ref{e15}) with respect to $R$, we obtain
\begin{eqnarray}
 \frac{d^{2}R}{d\phi^{2}}+R=\frac{3MR^{2}}{(1+2M\ell^{2}R^{3})^{2}}.
\label{e16}
\end{eqnarray}
We adopt the method of the previous section to investigate the influence of energy level $E^{2}$ on null geodesic structure. The behavior of effective potential with different values of parameters $\ell$ and $b$ is described in Fig. \ref{v11}.
\subsection{Circular geodesics}
\begin{figure}[H]
\centering
    \includegraphics[angle=0, width=0.37\textwidth]{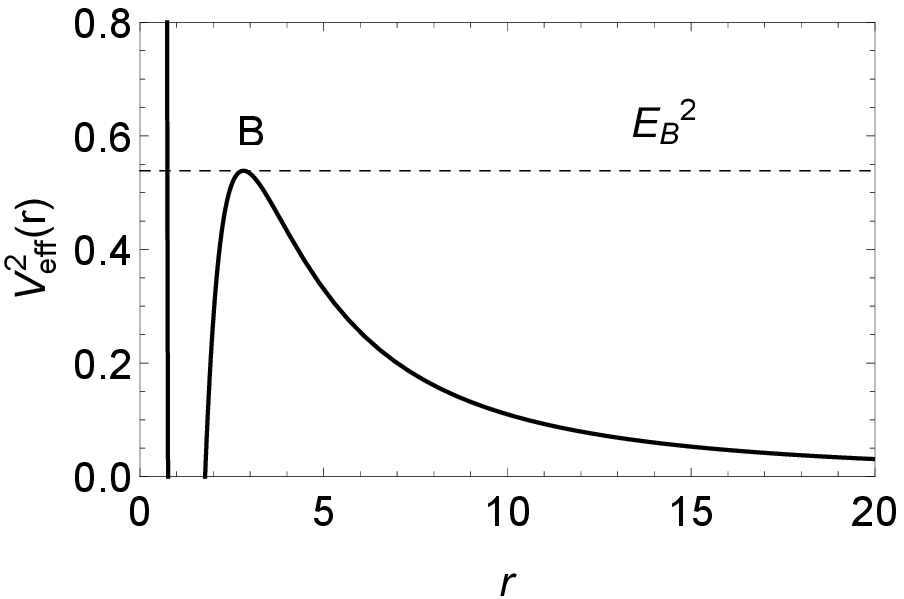}
     \\ \includegraphics[angle=0, width=0.3\textwidth]{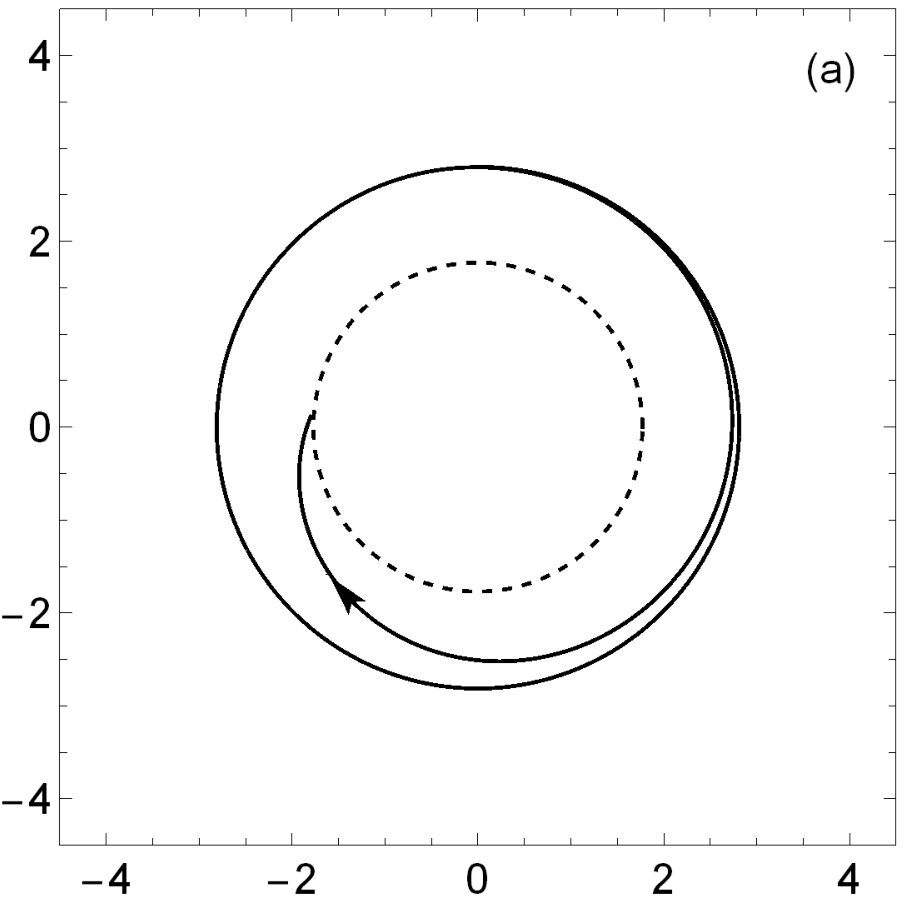}
     \ \ \ \ \includegraphics[angle=0, width=0.3\textwidth]{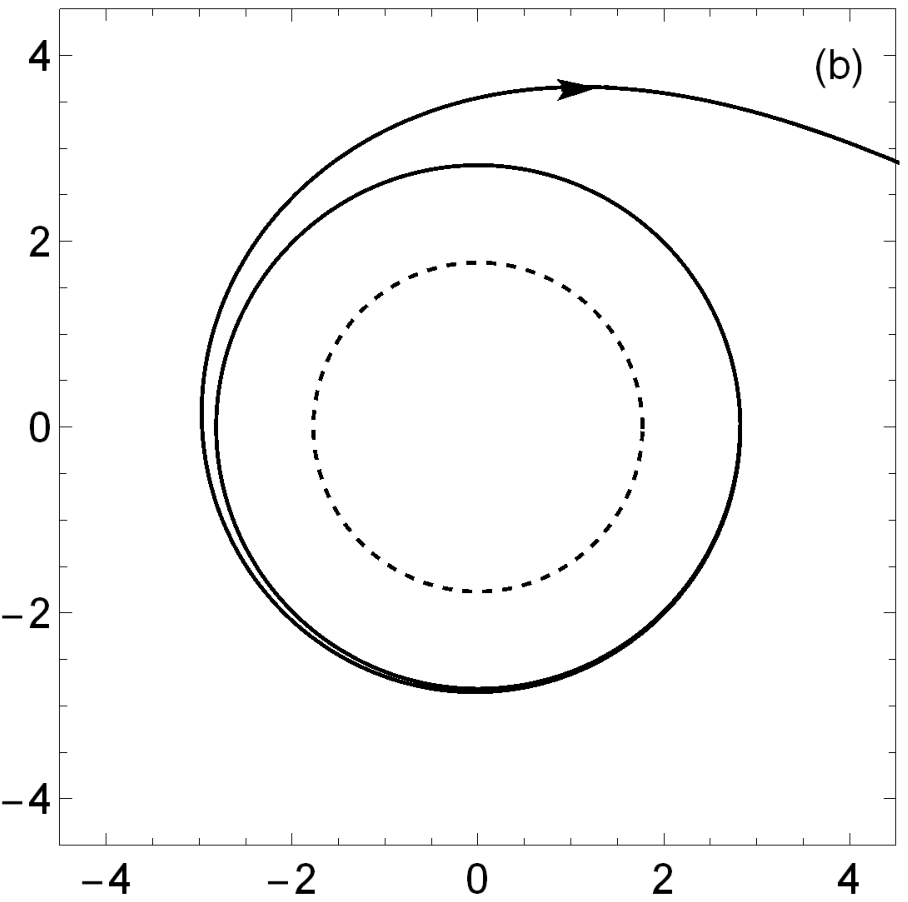}
     \vspace*{8pt}
\caption{Two kinds of unstable circular geodesics with $E^{2}$=0.54, $\ell$=0.60$M$, $b$=3.70$M$ and $M$=1 in the regular Hayward black hole space-time.\label{v12} }
\end{figure}

When the energy level $E^{2}=E^{2}_{B}$ the photons move on unstable circular orbits at $r$=$r_{B}$. On being perturbed, they may move from $r_{B}$ to infinity. Or, they may move from $r_{B}$ and fall into the black hole finally. The two cases are due to the initial conditions and outside perturbation which are shown in Fig. \ref{v12}.

\subsection{Escape geodesics}

\begin{figure}[H]
\centering
    \includegraphics[angle=0, width=0.4\textwidth]{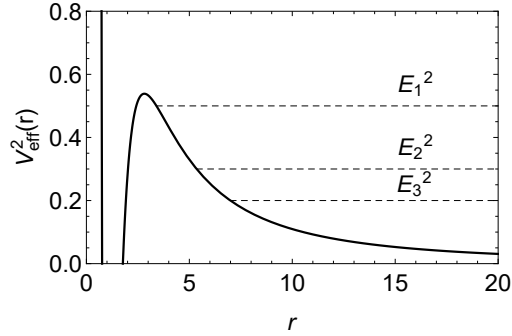}
     \vspace*{8pt}
\caption{The curve of effective potential with $\ell$=0.60$M$, $b$=3.70$M$ and $M$=1 in the regular Hayward black hole space-time. Here, $E^{2}_{1}$=$0.50$, $E^{2}_{2}$=$0.30$ and $E^{2}_{3}$=$0.20$. \label{v13} }
\end{figure}

\begin{figure}[H]
\centering
    \includegraphics[angle=0, width=0.3\textwidth]{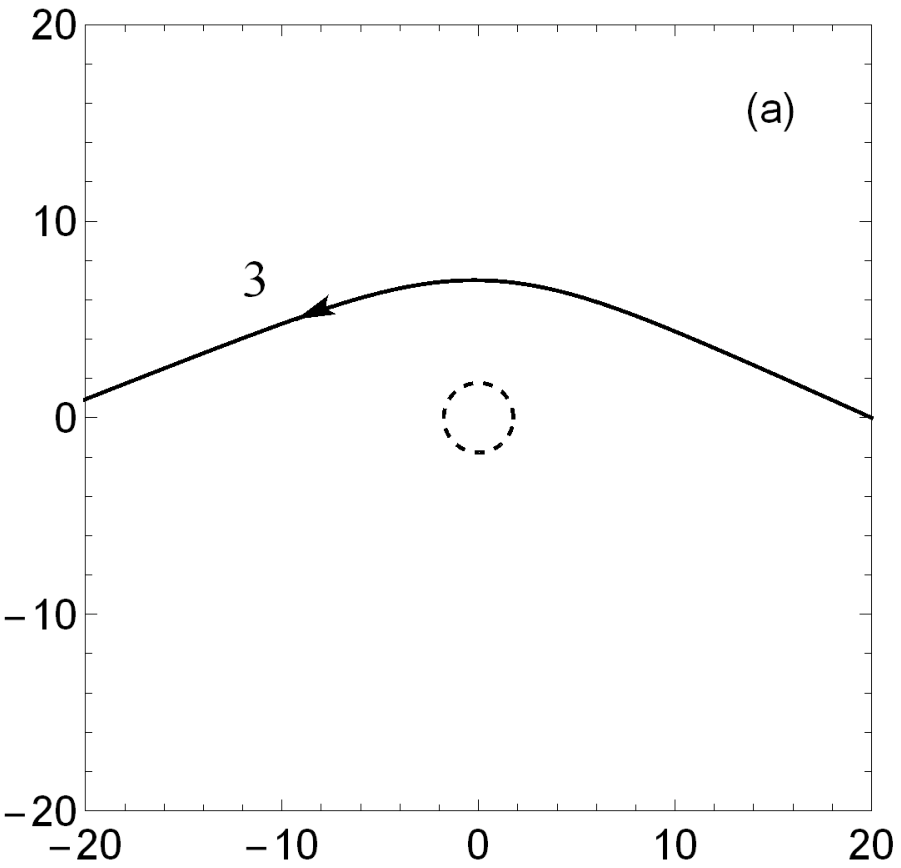}
      \ \ \ \ \includegraphics[angle=0, width=0.3\textwidth]{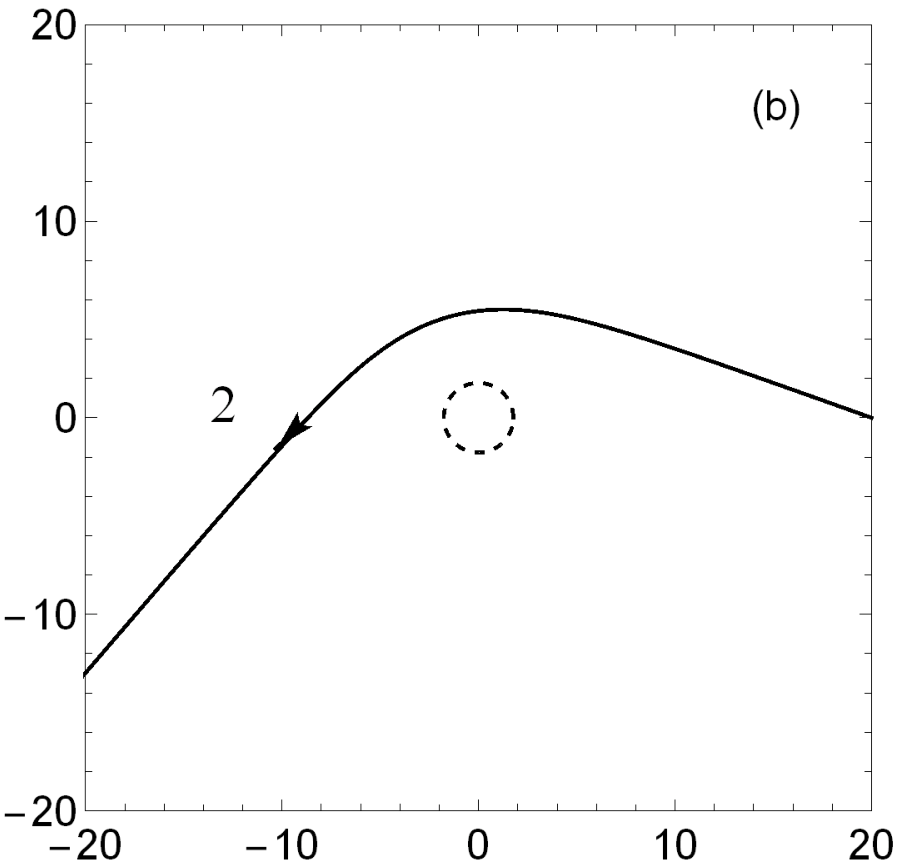}
     \\ \ \includegraphics[angle=0, width=0.3\textwidth]{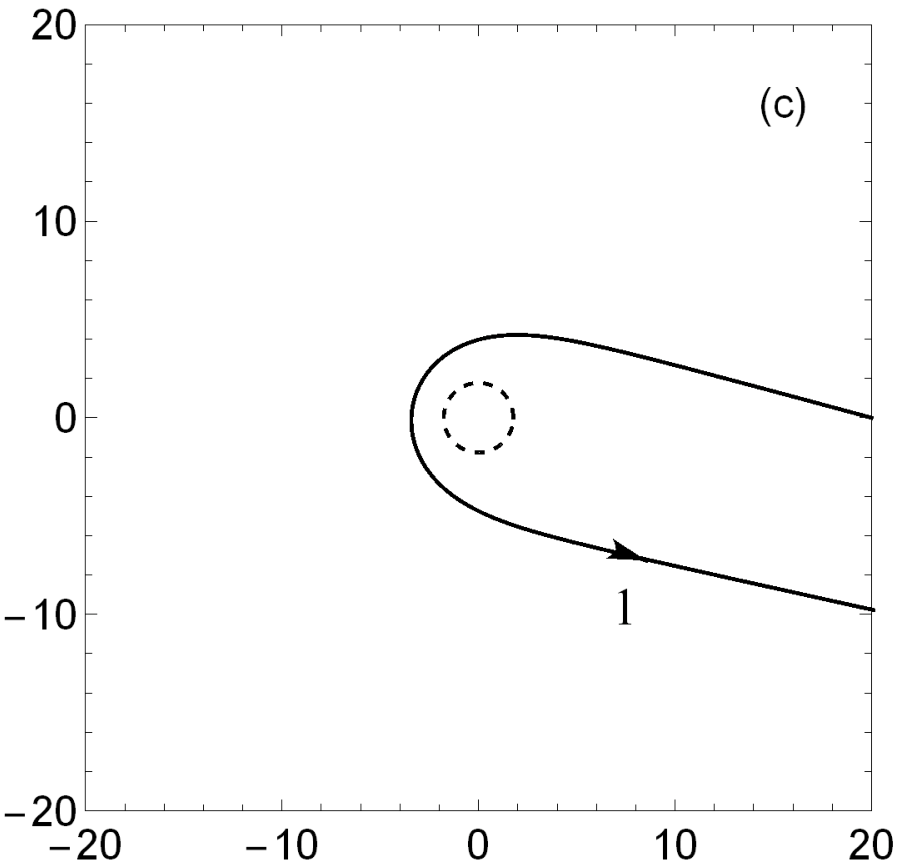}
      \ \ \ \  \includegraphics[angle=0, width=0.3\textwidth]{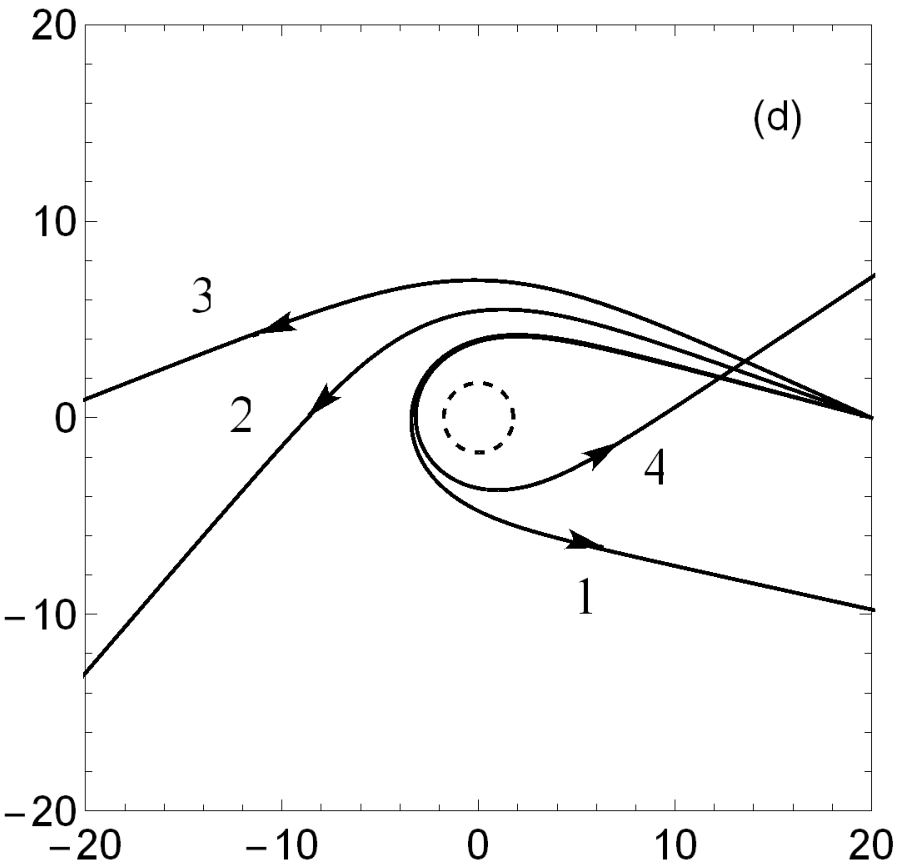}
     \vspace*{8pt}
\caption{The escape orbits with $\ell$=0.60$M$, $b$=3.70$M$, $M$=1 and different energy level ($E^{2}_{3}$=$0.20$, $E^{2}_{2}$=$0.30$, $E^{2}_{1}$=$0.50$ and $E^{2}_{4}>E^{2}_{1}$) in the regular Hayward black hole space-time. \label{v14} }
\end{figure}

Under $b=3.70M$ the escape orbits are shown in Fig. \ref{v14}. Fig. \ref{v14} (a), (b) and (c) correspond to the energy level $E^{2}_{1}$, $E^{2}_{2}$ and $E^{2}_{3}$(Fig. \ref{v13}), respectively. The influence of energy level on escape geodesics is investigated. It is found that the higher the energy, the higher the bending. When $E^{2}_{4}>E^{2}_{1}$, the orbits will intersect themselves.

\subsection{Absorbing geodesics}
\begin{figure}[H]
\centering
    \includegraphics[angle=0, width=0.36\textwidth]{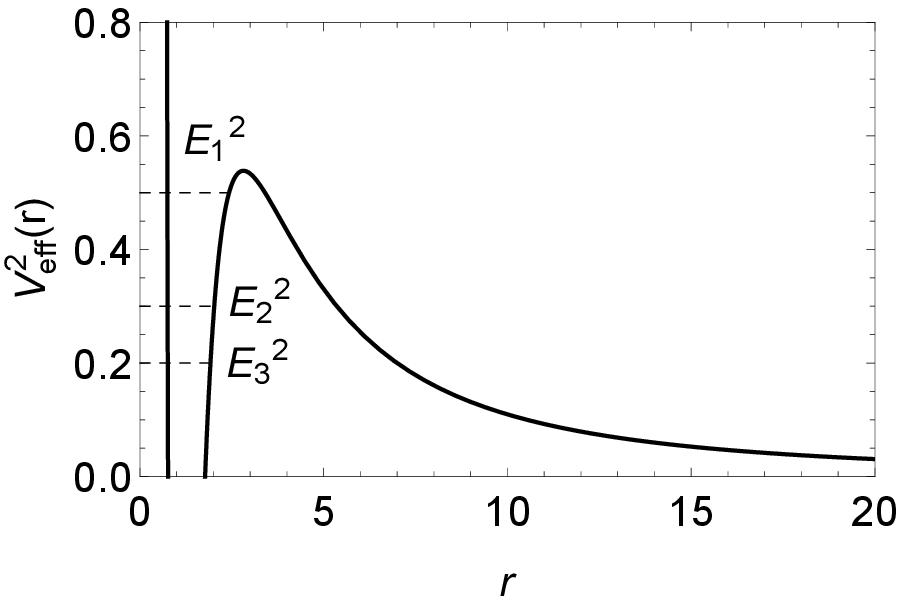}
      \ \ \ \ \includegraphics[angle=0, width=0.25\textwidth]{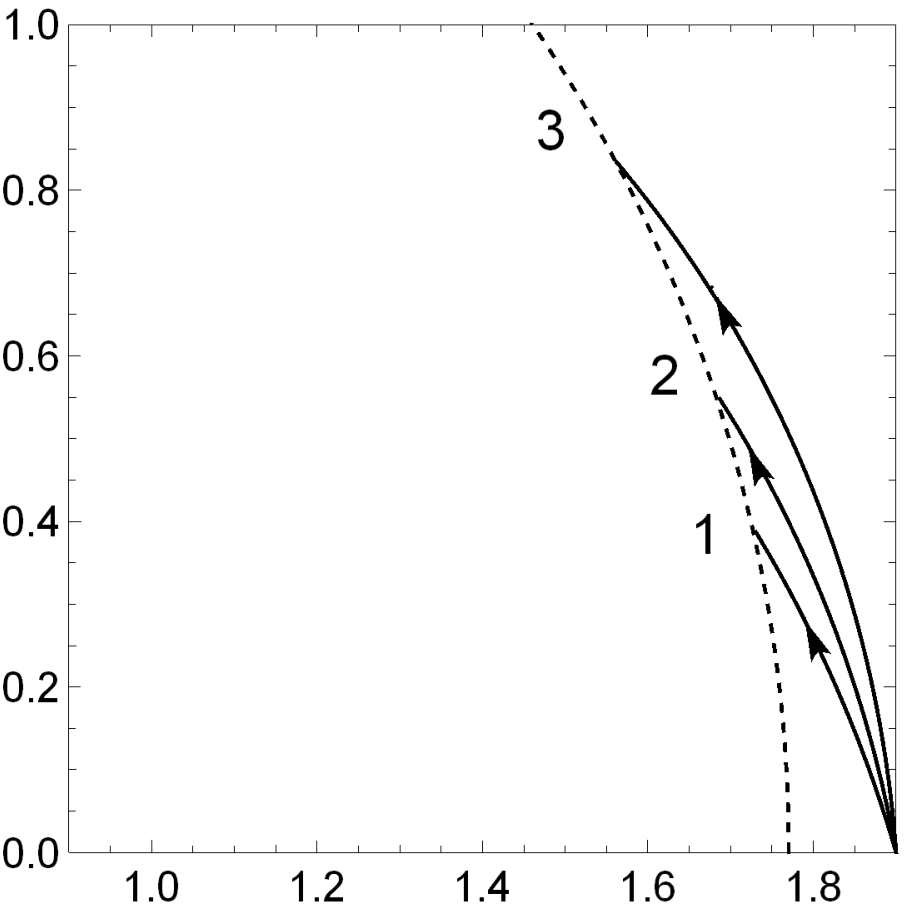}
     \vspace*{8pt}
\caption{The absorbing orbits with low energy level ($E^{2}_{1}$=$0.50$, $E^{2}_{2}$=$0.30$ and $E^{2}_{3}$=$0.20$) in the regular Hayward black hole space-time. \label{v15} }
\end{figure}

\begin{figure}[H]
\centering
     \ \includegraphics[angle=0, width=0.36\textwidth]{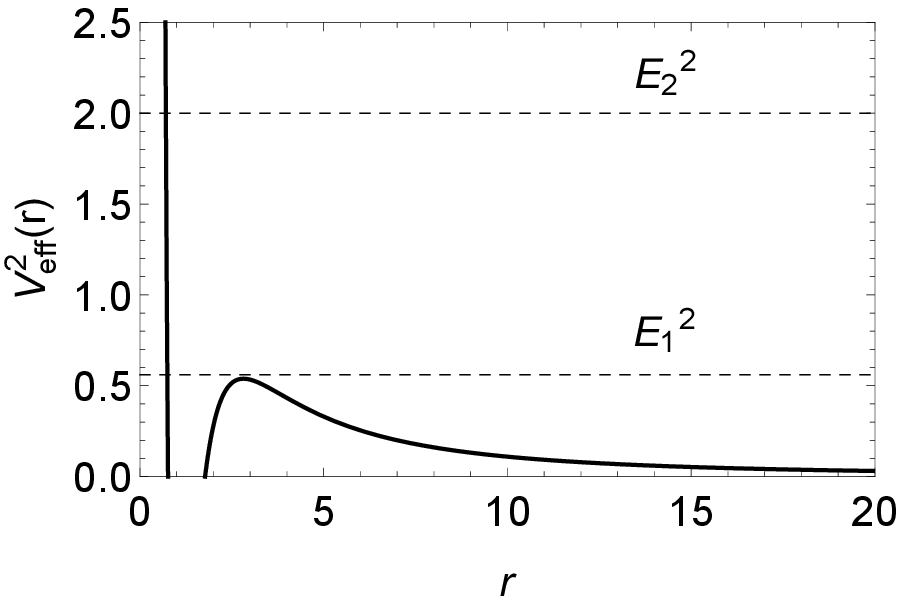}
      \ \ \ \   \includegraphics[angle=0, width=0.25\textwidth]{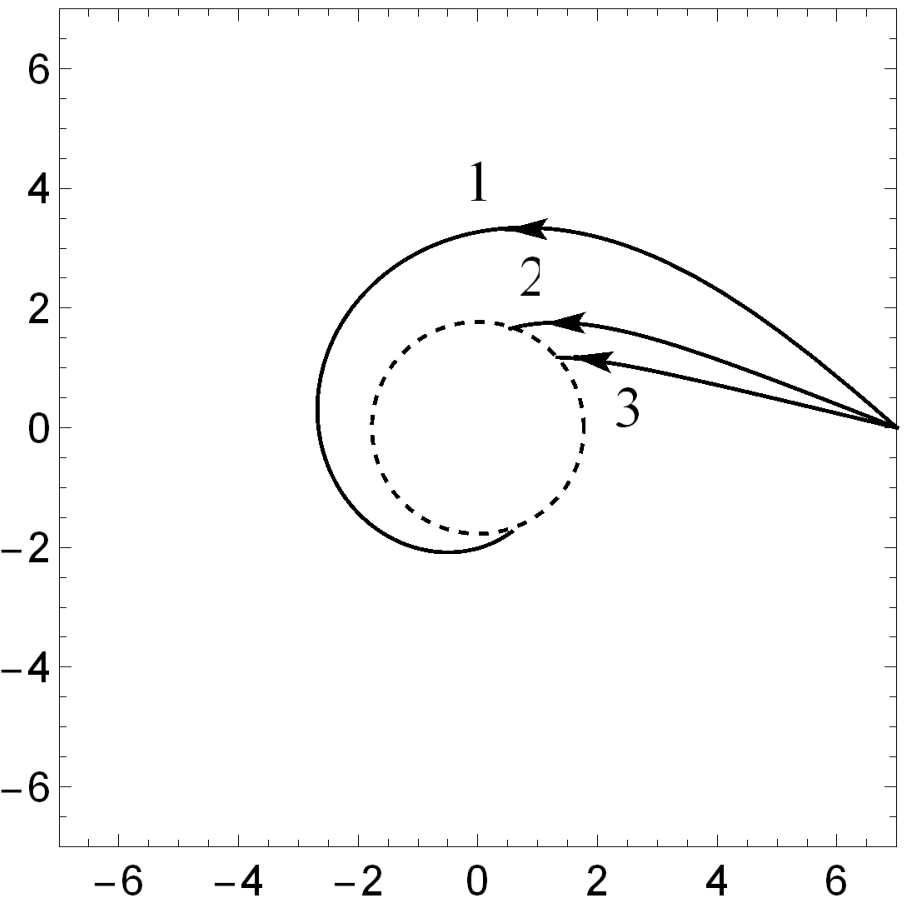}
     \vspace*{8pt}
\caption{The absorbing orbits with high energy level ($E^{2}_{1}$=$0.50$, $E^{2}_{2}$=$2.00$ and $E^{2}_{3}>E^{2}_{2}$) in the regular Hayward black hole space-time.  \label{v16} }
\end{figure}

We adopt the same energy level as Fig. \ref{v14} to plot the absorbing orbits, as shown in Fig. \ref{v15}. It can be seen that the higher the energy, the lower the bending. The test photons with higher energy level will fall into the black hole earlier.
From Fig. \ref{v16}, it is clear that the test photons with large enough energy level can also fall fast into the black hole when being thrown in the direction of the black hole.
\section{Conclusions}
By analyzing the lapse function, we obtain all possible types of horizons in the regular Hayward black hole space-time.
 Then, we give the equation of orbital motion and the effective potential, and distinguish four kinds of orbits (planetary orbits, escape orbits, circular orbits and absorbing orbits ) by analyzing the effective potential and solving the equation of orbital motion in the regular Hayward black hole space-time. The orbital stability and the precession direction of bound orbits are related to the angular momentum. The extreme values of angular momentum is $3.9512M$. For time-like geodesics, there exist a stable circular orbit, a bound orbit, two kinds of unstable circular orbits, a escape orbit and two kinds of absorbing orbits with $b>3.9512M$. When $b<3.9512M$, there are no escape orbits. Orbital types of the photons are simpler than those of the massive particles.  There are only two kinds of unstable circle orbits, one escape orbit and two kinds of absorbing orbits. We also find that the energy level can affect the curvature degree of the bound orbits, escape orbits and absorbing orbits.

{\bf Acknowledgments}
This work was supported in part by the National Natural Science Foundation of China (Grant No. 11565016), the Special Training Program for Distinguished Young Teachers of the Higher Education Institutions of Yunnan Province (Grant No. 1096837802), the Applied Basic Research Programs of Yunnan Provincial Science and Technology Department (Grant No. 2016FB008).


\begin{thebibliography}{}
\bibitem{hayward2006angular} Hayward, S. A.: Angular momentum conservation for dynamical black holes. Phys. Rev. D 74, 104013 (2006)
\bibitem{bardeen1968non} Bardeen, J. M.: Non-singular general-relativistic gravitational collapse. Proc. Int. Conf. GR5, Tbilisi 174, (1968)

\bibitem{zhou2012geodesic} Zhou, S., Chen, J. H., Wang, Y. J.: Geodesic Structure of Test Particle in Bardeen Spacetime. Int. J. Mod. Phys. D 21, 1250077 (2012)

 \bibitem{lin2013quasinormal} Lin, K. Li, J., Yang, S. Z.: Quasinormal modes of Hayward regular black hole. Int. J. Thero. Phys. 52, 3771-3778 (2013)
\bibitem{abbas2014geodesic} Abbas, G., Sabiullah, U.: Geodesic study of regular Hayward black hole. Astrophys. Space Sci. 352, 769-774 (2014)
\bibitem{amir2015rotating} Amir, M., Ghosh, S. G.: Rotating Hayward's regular black hole as particle accelerator. JHEP 1507, 015 (2015)
\bibitem{amir2016collision} Amir, M.: Collision of two general particles around a rotating regular Hayward's black holes. Eur. Phys. J. C. 76, 532 (2016)


\bibitem{bozza2004time} Bozza V., Mancini L.: Time delay in black hole gravitational lensing as a distance estimator. Gen. Relativ. Gravit. 36, 435-450 (2004)


\bibitem{delorenci2001dyadosphere} De Lorenci, V. A., Figueiredo, N., Fliche, H. H., Novello, M.: Dyadosphere bending of light. Astron. Astrophys. 369, 690-693 (2001)

\bibitem{virbhadra2008relativistic} Virbhadra, K. S.: Relativistic images of Schwarzschild black hole lensing. Phys. Rev. D 79, 083004 (2009)

\bibitem{ross1971gravitational} Ross, D. K.: Gravitational red-shift. Il Nuovo Cimento B 2, 55-62 (1971)



\bibitem{greenberg1981apsidal} Greenberg, R.: Apsidal precession of orbits about an oblate planet. Astronom. J. 86, 912-916 (1981)



\bibitem{cardoso2003quasinormal} Cardoso, V., Konoplya, R., Lemos, J. P. S.: Quasinormal frequencies of Schwarzschild black holes in anti-de Sitter spacetimes: A complete study of the overtone asymptotic behavior. Phys. Rev. D 68, 044024 (2003)

\bibitem{berti20009quasinormal} Berti, E., Cardoso, V., Starinets, A. O.: Quasinormal modes of black holes and black branes. Classical Quant. Grav. 26, 163001 (2009)


\bibitem{konoplya2011quasinormal} Konoplya, R. A., Zhidenko, A.: Quasinormal modes of black holes: from astrophysics to string theory. Rev. Mod. Phys. 83, 793 (2011)




\bibitem{chandrasekhar1984the} Chandrasekhar, S., Thorne K. S.: The mathematical theory of black holes. Phys. Today 37, 5-26 (1984)

\bibitem{stuchlik1991null} Stuchlik, Z., Calvani, M.: Null geodesics in black hole metrics with non-zero cosmological constant. Gen. Relativ. Gravit. 23, 507-519 (1991)


\bibitem{cruz1994geodesic} Cruz, N., Martinez, C., Pena, L.: Geodesic Structure of the (2 + 1)-dimensional BTZ black hole. Classical Quant. Grav. 11, 2731-2739 (1994)


\bibitem{beem1997stability} Beem, J. K.: Stability of geodesic structures. Nonlinear Anal. Theor. 30, 567-570 (1997)
\bibitem{podolsky1999thestructure} Podolsky, J.: The structure of the extreme Schwarzschild-de Sitter space-time. Gen. Relativ. Gravit. 31, 1703-1725 (1999)

\bibitem{breton2002geodesic} Bret\'{o}n, N.: Geodesic structure of the Born-Infeld black hole. Classical Quant. Grav. 19, 601-612 (2002)


\bibitem{allison2003geodesic} Allison, D., Unal B.: Geodesic structure of standard static space-times. J. Geom. Phys. 46, 193-200 (2003)
\bibitem{kraniotis2004precise} Kraniotis, G. V.: Precise relativistic orbits in Kerr and Kerr-(anti) de Sitter spacetimes. Class. Quant. Grav. 21 4743 (2004)
\bibitem{stuchlk2004equatorial} Stuchlak, Z., Slan\'{y}, P.: Equatorial circular orbits in the Kerr-de Sitter spacetimes. Phys. Rev. D 69, 064001 (2004)

\bibitem{hackmann2008geodesic} Hackmann, E., L{\"a}mmerzahl, C.: Geodesic equation in Schwarzschild-(anti-) de Sitter space-times: Analytical solutions and applications. Phys. Rev. D 78, 024035 (2008)
\bibitem{cardoso2009geodesic} Cardoso, V., Miranda, A. S., Berti, E., Witek, H., Zanchin, V. T.: Geodesic stability, Lyapunov exponents, and quasinormal modes. Phys. Rev. D 79, 064016 (2009)

\bibitem{abdujabbarov2010test} Abdujabbarov, A., Ahmedov, B.: Test particle motion around a black hole in a braneworld. Phys. Rev. D 81, 044022 (2010)
\bibitem{muller2011studying} Muller, T., Frauendiener, J.: Studying null and time-like geodesics in the classroom. Eur. J. Phys. 32, 747-759 (2011)

\bibitem{halilsoy2013rindler} Halilsoy, M., Gurtug, O., Mazharimousavi, S. H.: Rindler modified Schwarzschild geodesics. Gen. Relativ. Gravit. 45, 2363-2381 (2013)
\bibitem{chakraborty2014inner-most} Chakraborty, C.: Inner-most stable circular orbits in extremal and non-extremal Kerr-Taub-NUT spacetimes. Eur. Phys. J. C 74, 2759 (2014)

\bibitem{pradhan2015circular} Pradhan, P.: Circular geodesics in the Kerr-Newman-Taub-NUT spacetime. Class. Quant. Grav. 32, 165001 (2015)

\bibitem{zhang2015time-like} Zhang, R. J., Zhou, S., Chen, J. H.: Time-like geodesic structure in massive gravity. Gen. Relativ. Gravit. 47, 128 (2015)

\bibitem{chandler2015geodesic} Chandler, J., Emam, M. H.: Geodesic structure of five-dimensional nonasymptotically flat 2-branes. Phys. Rev. D 91, 125024 (2015)


\bibitem{konoplya2017areeikonal} Konoplya, R. A., Stuchlak, Z.: Are eikonal quasinormal modes linked to the unstable circular null geodesics?. Phys Lett. B 771, 597-602 (2017)
\bibitem{farrugia2017thermodynamic} Farrugia, C., Sultana, J.: Thermodynamic geodesics of a Reissner Nordstr{\"o}m black hole. Gen. Relativ. Gravit. 49, 4 (2017)

\bibitem{azam2017geodesic} Azam, M., Abbas, G., Sumera, S., Nizami, A. R.: Geodesic structure of magnetically charged regular black hole. Int. J. Geom. Methods Mod. Phys. 14, 1750120 (2017)

\bibitem{azam2017geodesic2} Azam, M., Abbas, G., Sumera, S.: Geodesic Motion Around Regular Magnetic Black Hole in Non-
 minimal Einstein-Yang-Mills Theory. Can. J. Phys. 95, 1062-1067 (2017)

\bibitem{chen2010timelike} Chen, J. H., Wang, Y. J.: Timelike Geodesic Motion in Horava-Lifshitz Spacetime. Int. J. Mod. Phys. 25, 1439-1448 (2010)

\bibitem{sheng2011time-like} Zhou, S, Chen, J. H., Wang, Y. J.: Time-like geodesic structure of a spherically symmetric black hole in the brane-world. Chinese Physics B, 20, 100401-100401 (2011)

\bibitem{li2014particle} Li, E. K., Zhang, Y.: Particle motion in the Schwarzschild-Quintessence space-time. Astrophys. Space Sci. 350, 361-366 (2014)
\bibitem{dean1999phase} Dean, B.: Phase-plane analysis of perihelion precession and Schwarzschild orbital dynamics. Am. J. Phys. 67, 78-86 (1999)
\bibitem{zhang2014orbital} Zhang, Y., Geng, J. L., Li, E. K.: Orbital dynamics of the gravitational field of stringy black holes. Mod. Phys. Lett. A 29, 1450144 (2014)
\bibitem{zhang2014orbital2} Zhang, Y., Li, E. K., Geng, J. L.: Orbital dynamics of the gravitational field in Bardeen space-time. Astrophys. Space Sci. 351, 665-669 (2014)
\bibitem{yi2006geodesics} Zeng, Y., Lu, J. L., Wang, Y. J.: Geodesics of Spherical Dilaton Spacetimes. Chinese Phys. Lett. 23, 1648-1651 (2006)

\bibitem{Petr2009Quantum} Ho\v{r}ava, P.: Quantum gravity at a Lifshitz point. Phys. Rev. D, 79, 084008 (2009)

\bibitem{heydarifard2009spherically} Heydarifard, M., Sepangi, H. R.: Spherically symmetric solutions and gravitational collapse in brane-worlds. J. Cosmol. Astropart. P. 2009, 029-029 (2009)


\bibitem{frolov1987charged} Frolov, V. P., Zelnikov, A. I., Bleyer, U.: Charged Rotating Black Hole from Five Dimensional Point of View.  Ann. Phys. Berlin. 499, 371-377 (1987)



\end{thebibliography}
\end{document}